\documentclass[11pt]{article}
\usepackage{amsfonts,amsmath}
\usepackage{amssymb}
\usepackage{fancyhdr}
\usepackage{slashed}
\usepackage{graphicx}
\usepackage{subfigure}
\usepackage{color}

\usepackage{hyperref}

\usepackage[titletoc]{appendix}

\def\hybrid{\topmargin -20pt    \oddsidemargin 0pt
        \headheight 0pt \headsep 0pt
        \textwidth 6.25in       
        \textheight 9.25in       
        \marginparwidth .875in
        \parskip 5pt plus 1pt   \jot = 1.5ex}

\hybrid

\def\baselinestretch{1.2}

\catcode`\@=11

\def\marginnote#1{}
\newcount\hour
\newcount\minute
\newtoks\amorpm
\hour=\time\divide\hour by60
\minute=\time{\multiply\hour by60 \global\advance\minute by-\hour}
\edef\standardtime{{\ifnum\hour<12 \global\amorpm={am}%
        \else\global\amorpm={pm}\advance\hour by-12 \fi
        \ifnum\hour=0 \hour=12 \fi
        \number\hour:\ifnum\minute<10 0\fi\number\minute\the\amorpm}}
\edef\militarytime{\number\hour:\ifnum\minute<10 0\fi\number\minute}

\def\draftlabel#1{{\@bsphack\if@filesw {\let\thepage\relax
   \xdef\@gtempa{\write\@auxout{\string
      \newlabel{#1}{{\@currentlabel}{\thepage}}}}}\@gtempa
   \if@nobreak \ifvmode\nobreak\fi\fi\fi\@esphack}
        \gdef\@eqnlabel{#1}}
\def\@eqnlabel{}
\def\@vacuum{}
\def\draftmarginnote#1{\marginpar{\raggedright\scriptsize\tt#1}}

\def\draft{\oddsidemargin -.5truein
        \def\@oddfoot{\sl preliminary draft \hfil
        \rm\thepage\hfil\sl\today\quad\militarytime}
        \let\@evenfoot\@oddfoot \overfullrule 3pt
        \let\label=\draftlabel
        \let\marginnote=\draftmarginnote
   \def\@eqnnum{(\theequation)\rlap{\kern\marginparsep\tt\@eqnlabel}%
\global\let\@eqnlabel\@vacuum}  }

\def\preprint{\twocolumn\sloppy\flushbottom\parindent 2em
        \leftmargini 2em\leftmarginv .5em\leftmarginvi .5em
        \oddsidemargin -.5in    \evensidemargin -.5in
        \columnsep .4in \footheight 0pt
        \textwidth 10.in        \topmargin  -.4in
        \headheight 12pt \topskip .4in
        \textheight 6.9in \footskip 0pt
        \def\@oddhead{\thepage\hfil\addtocounter{page}{1}\thepage}
        \let\@evenhead\@oddhead \def\@oddfoot{} \def\@evenfoot{} }

\def\numberbysection{\@addtoreset{equation}{section}
        \def\theequation{\thesection.\arabic{equation}}}

\def\underline#1{\relax\ifmmode\@@underline#1\else
        $\@@underline{\hbox{#1}}$\relax\fi}

\def\titlepage{\@restonecolfalse\if@twocolumn\@restonecoltrue\onecolumn
     \else \newpage \fi \thispagestyle{empty}\c@page\z@
        \def\thefootnote{\fnsymbol{footnote}} }

\def\endtitlepage{\if@restonecol\twocolumn \else \newpage \fi
        \def\thefootnote{\arabic{footnote}}
        \setcounter{footnote}{0}}  

\catcode`@=12
\relax

\def\figcap{\section*{Figure Captions\markboth
        {FIGURECAPTIONS}{FIGURECAPTIONS}}\list
        {Figure \arabic{enumi}:\hfill}{\settowidth\labelwidth{Figure
999:}
        \leftmargin\labelwidth
        \advance\leftmargin\labelsep\usecounter{enumi}}}
 \relax
\def\tablecap{\section*{Table Captions\markboth
        {TABLECAPTIONS}{TABLECAPTIONS}}\list
        {Table \arabic{enumi}:\hfill}{\settowidth\labelwidth{Table
999:}
        \leftmargin\labelwidth
        \advance\leftmargin\labelsep\usecounter{enumi}}}
 \relax
\def\reflist{\section*{References\markboth
        {REFLIST}{REFLIST}}\list
        {[\arabic{enumi}]\hfill}{\settowidth\labelwidth{[999]}
        \leftmargin\labelwidth
        \advance\leftmargin\labelsep\usecounter{enumi}}}
 \relax
\makeatletter
\newcounter{pubctr}
\def\publist{\@ifnextchar[{\@publist}{\@@publist}}
\def\@publist[#1]{\list
        {[\arabic{pubctr}]\hfill}{\settowidth\labelwidth{[999]}
        \leftmargin\labelwidth
        \advance\leftmargin\labelsep
        \@nmbrlisttrue\def\@listctr{pubctr}
        \setcounter{pubctr}{#1}\addtocounter{pubctr}{-1}}}
\def\@@publist{\list
        {[\arabic{pubctr}]\hfill}{\settowidth\labelwidth{[999]}
        \leftmargin\labelwidth
        \advance\leftmargin\labelsep
        \@nmbrlisttrue\def\@listctr{pubctr}}}
 \relax
\makeatother
\newskip\humongous \humongous=0pt plus 1000pt minus 1000pt

\newif\ifdtup

\relax

\def\be{\begin{equation}}
\def\ee{\end{equation}}
\def\ba{\begin{eqnarray}}
\def\ea{\end{eqnarray}}

\def\del{\partial}

\def\k{\kappa}
\def\r{\rho}
\def\a{\alpha}

\def\e{\epsilon}

\def\th{\theta}

\def\m{\mu}

\def\l{\lambda}

\def\s{\sigma}

\def\vphi{\varphi}
\def\cN{{\cal N}}

\def\cL{{\cal L}}

  \def\cL{{\cal L}}
 \def\cN{{\cal N}} 
  
\def\cS{{\cal S}}

\newcommand{\prt}[1]{{\left( {#1} \right)}}

\def\no{\noindent}

\def\delb{\overline\del}
\def\IR{\relax{\rm I\kern-.18em R}}

\def\IR{\relax{\rm I\kern-.18em R}}
\def\IL{\relax{\rm I\kern-.18em L}}

\def\inv{^{\raise.15ex\hbox{${\scriptscriptstyle -}$}\kern-.05em 1}}

\def\cL{{\cal L}}

\def\bea{\begin{eqnarray}}
\def\eea{\end{eqnarray}}
\newcommand{\eq}[1]{(\ref{#1})}
\def\nn{\nonumber}
\def\del{\partial}
\newcommand{\la}[1]{\label{#1}}

\def\a{\alpha}      \def\da{{\dot\alpha}}

\def\e{\epsilon}

\def\k{\kappa}
\def\l{\lambda} 
\def\m{\mu} 
\def\o{\omega}
 
\def\r{\rho}
\def\s{\sigma}  
\def\t{\tau}
\def\th{\theta}

\def \da {\dot{\alpha}}

\def \dt {\dot{t}}

\def \dr {\dot{r}}

\def \dR {\dot{R}}

\def \dth {\dot{\theta}}

\definecolor{markcolor2}{rgb}{1,0,0}

\definecolor{markcolor3}{rgb}{0,1,0}

\newcommand{\half}{\frac{1}{2}}
\newcommand{\CPone}{\mathbb{C} \mathrm{P}^{1}}
\newcommand{\CPtwo}{\mathbb{C} \mathrm{P}^{2}}
\newcommand{\CPthree}{\mathbb{C} \mathrm{P}^{3}}
\newcommand{\RPthree}{\mathbb{R} \mathrm{P}^{3}}
\newcommand{\Srm}{\mathrm{S}}
\newcommand{\AdS}{\mathrm{AdS}}
\newcommand{\diff}{\mathrm{d}}
\newcommand{\jbar}{{\bar{\jmath}}}
\newcommand{\zb}{\bar{z}}
\newcommand{\wb}{\bar{w}}

\newcommand{\Acal}{\mathcal{A}}

\newcommand{\Hcal}{\mathcal{H}}
\newcommand{\Ncal}{\mathcal{N}}
\newcommand{\Scal}{\mathcal{S}}
\newcommand{\Lcal}{\mathcal{L}}

\newcommand{\SU}{\mathrm{SU}}
\newcommand{\SO}{\mathrm{SO}}

\newcommand{\Sp}{\mathrm{Sp}}
\newcommand{\Urm}{\mathrm{U}}

\newcommand{\alphadot}{{\dot{\alpha}}}
\newcommand{\thetadot}{{\dot{\theta}}}
\newcommand{\alphadotdot}{{\ddot{\alpha}}}
\newcommand{\thetadotdot}{{\ddot{\theta}}}

\begin{document}

\renewcommand{\theequation}{\thesection.\arabic{equation}}
\csname @addtoreset\endcsname{equation}{section}

\newcommand{\beq}{\begin{equation}}
\newcommand{\eeq}[1]{\label{#1}\end{equation}}
\newcommand{\ber}{\begin{eqnarray}}
\newcommand{\eer}[1]{\label{#1}\end{eqnarray}}
\newcommand{\eqn}[1]{(\ref{#1})}
\begin{titlepage}

\begin{center}

~
\vskip 1 cm

{\Large
\bf Non-integrability and Chaos with Unquenched Flavor}

\vskip 0.5in

{\bf Dimitrios Giataganas$^{1}$ \phantom{x}and\phantom{x}Konstantinos Zoubos}$^{2,3}$
\vskip 0.1in
{\em
  ${}^1$  Physics Division, National Center for Theoretical
  Sciences, \\
 National Tsing-Hua University, Hsinchu, 30013, Taiwan

\vskip .15in
${}^2$
Department of Physics, University of Pretoria\\
Private Bag X20, Hatfield 0028, South Africa
\vskip .15in
${}^3$
National Institute for Theoretical Physics (NITheP) \\
Gauteng, South Africa
\\\vskip .1in
{\tt dimitrios.giataganas@cts.nthu.edu.tw , kzoubos@up.ac.za}\\
}

\vskip .2in
\end{center}

\vskip .4in

\centerline{\bf Abstract}

We study (non-)integrability and the presence of chaos in gravity dual backgrounds of strongly coupled gauge theories with
unquenched flavor, specifically of the four-dimensional $\cN=2$ super Yang-Mills theory and the three-dimensional  ABJM theory. By
examining  string motion on the geometries corresponding to backreacted D3/D7 and D2/D6 systems, we show that integrable
theories with quenched flavor become non-integrable when the virtual quark loops are taken into account. For the string
solutions in the backreacted D3/D7 system, we compute the leading Lyapunov exponent which turns out to saturate to a
positive value as the number of flavors increases. The exponent depends very weakly on the number of flavors when they
approach the number of colors. This suggests that once a particular flavor number in the theory is reached, a further
increase does not lead to more severe chaotic phenomena, implying certain saturation effects on chaos.
\no
\end{titlepage}
\vfill
\eject


\noindent


\def\baselinestretch{1.2}
\baselineskip 19 pt
\noindent


\setcounter{equation}{0}

\tableofcontents
\section{Introduction}

The study of integrability (see e.g. \cite{Beisert:2010jr, Bombardelli:2016rwb} for reviews) has proved to be a
very useful direction in obtaining exact results in AdS/CFT. An important question concerns which gauge theory
systems admit integrability or not. A fruitful approach to this problem from the dual gravity side has been to
apply techniques of analytic non-integrability. The technique, introduced in \cite{Basu:2011fw}, consists of
finding solutions for classical string motion on the gravitational background dual to a given gauge theory
which can be reduced to a Hamiltonian system on which one can apply the variational techniques of analytic
non-integrability outlined in \cite{MoralesRuiz99}.

Moreover recently there has been an extensive discussion of chaos in quantum field theories, which has further
potential applications relating the physics of black holes and quantum information. The Lyapunov exponent $\l$ of
out-of-time-ordered correlators has been proved to be bounded by the temperature $T$ of the theory as
$\l\le 2 \pi T/\hbar$ \cite{Maldacena:2015waa} using shock waves near black hole horizons
\cite{Shenker:2013pqa,Shenker:2013yza}, or by looking at the particle motion near the horizon \cite{Hashimoto:2016dfz}.

In this work we aim to study, from the dual gravity side, the integrability of theories with unquenched flavor
(see for example \cite{Erdmenger:2007cm,Nunez:2010sf} and references therein), where the virtual quark loops are
taken into account contributing to the gauge propagators. Moreover we discuss in detail chaotic string motion on
these backgrounds and, at least in the limit we are working with, we find a convergent Lyapunov exponent to a
non-zero value, although no black hole horizon is present.

We initiate our study in the case of four-dimensional theories with $\Ncal=2$ supersymmetry,  obtained by adding
D7-branes to a D3-brane system as in \cite{Karch:2002sh}, and taking into account their backreaction.
The background for this system was constructed in \cite{Burrington:2004id,Kirsch:2005uy}. We will work in the ``near-core'' regime of \cite{Kirsch:2005uy} where the solution is known analytically.  In the unquenched case the effects of creation and annihilation of virtual quark-anti-quark pairs on the gauge degrees of freedom are taken into account. On the gravity side, this is translated as having the Dp-branes backreacting on the ``gluonic'' pure AdS space. In principle this happens when the number $N_f$ of flavor branes approaches the number $N_c$ of the color branes. In this limit the theory located at the D3/D7-brane intersection has positive beta function and possesses a chiral anomaly. Therefore the gravity dual background ceases to be conformal as can be naturally found by taking into account the effects of backreaction, while the axion and the dilaton become non-trivial. The closed string solutions we find are exactly in the region where the approximation closely follows the exact numerical solution.

We perturb the closed string solutions and apply variational methods to obtain the Normal Variational Equation (NVE) and show that it does not have Liouvillian solutions.  Therefore we show that string motion is  non-integrable on this background. The integrability of our solutions is restored for particle motion or when we switch off the backreaction of the D7-flavor branes. We moreover examine chaotic string motion in the non-integrable theory. We find, for the string solutions we examine, that the Lyapunov exponent in the Veneziano regime is surprisingly saturated to a value and depends weakly on the number of flavors. The value depends on the energy density of the system and its other parameters.  By pushing the backreacted solution to the limits of its validity by reducing the number of flavors, we see that the maximum value of the Lyapunov exponent is obtained very quickly as we increase the flavors. We note that our analysis is done at zero temperature and the gravity dual theory has no black hole horizon. Furthermore, looking at the more generic features of the backreacted Dp-Dq brane geometries, we expect that integrability is broken in the unquenched flavor limit for a wider class of field theories.

Then we move on to examine the case of adding unquenched flavor to the ABJM model  \cite{Aharony:2008ug}.  This solution is obtained by introducing D6-branes and taking into account their backreaction. These backgrounds were constructed in \cite{Conde:2011sw} for the case with massless flavors, and extended to the massive case in \cite{Bea:2013jxa}. Unlike the D3/D7 case, these backgrounds are exact, while the D6 branes are smeared along the internal space. After reviewing the solution of \cite{Conde:2011sw}, we consider the general classical string equations of motion on this background and truncate them to a consistent ansatz which is suitable for applying analytic non-integrability techniques. We find analytically closed string solutions and we derive the NVE to study its solutions using the Kovacic algorithm.  We conclude that, as expected and in agreement with the  above D3/D7 analysis, the backreacted flavor deformation breaks the integrability present in the ABJM case. However we also point out a puzzling case where for a single string ansatz we do not find Liouvillian solutions of the corresponding NVE even when the backreaction is switched off -- conflicting our expectations of integrability in that limit. By investigating in depth the geometry we argue that it should be an issue of the string solution in the particular coordinate system used to obtain the backreaction.

As a byproduct of our study of the unquenched ABJM case, we study integrability in a range of other theories. There are special values of the parameter that controls the squashing of the space and the number of flavors, where the background is dual to other theories as well. For example, these backgrounds also appear as the IR fixed points of D2-D6 brane system flows. By computing the backreaction of the $N_f$ flavor D6-branes smeared over a six-dimensional nearly K\"{a}hler manifold, to the $N_c$ color branes, it has been found that the solutions flow to an $\AdS_4$ fixed point dual to Chern-Simons matter theory and which is the special case of the metric we present in section 4 \cite{Faedo:2015ula}. The family of squashed $\CPthree$ metrics is relevant also for the construction of squashed seven-sphere metrics, which are $\Srm^1$ bundles over a squashed $\CPthree$ base.  The special case $q=5$ corresponds to an $\cN=1$ supergravity background, being a gravity dual of superconformal Chern-Simons matter theory with $SO\prt{5}\times U\prt{1}$ global symmetry \cite{Ooguri:2008dk}. Therefore, our work applies in certain cases for different types of theories.

There are several works where non-integrability and/or chaos has been studied in an AdS/CFT context.  In \cite{Hashimoto:2016wme} the time evolution of the homogeneous quark condensate in supersymmetric $\cN=2$ QCD was studied and it was found, by looking at probe D-branes, that there exists an energy density where the phase space is dominated by chaos.  Methods similar to analytic non-integrability have been applied to theories such as $\AdS_5\times T^{pq}$ \cite{Basu:2011fw}, Dp-brane backgrounds \cite{Stepanchuk:2012xi}, the Lunin-Maldacena background dual to the $\beta$-deformations of $\Ncal=4$ SYM \cite{Giataganas:2013dha},  confining backgrounds \cite{Basu:2012ae,Ishii:2016rlk},  in theories beyond the planar limit \cite{Chervonyi:2013eja} although at special large N limits some integrability does appear \cite{Koch:2011hb}, and in non-relativistic theories \cite{Giataganas:2014hma}. Other works along these directions include \cite{ Farahi:2014lta,Ma:2014aha,Asano:2015eha, Asano:2015qwa,Panigrahi:2016zny, Asano:2016qsv,  Basu:2016zkr}.

Our paper is organised as follows. In section 2, we introduce a very compact notation to analyse the string equations of motion for a generic metric. In section 3, we study the integrability of the $\cN=2$ super Yang-Mills theory with $N_f$ hypermultiplets in the fundamental representation and one in the adjoint representation of the $SU(N_c)$ gauge group. We present the string solutions, the variational analysis and the proof of non-integrability. Moreover we examine the chaotic motion by solving the string equations numerically and we extract the Lyapunov exponent. In the next section 4, we study the geometry of the ABJM background with backreaction. We present several sting solutions and the NVE showing non-integrability. Moreover we mention a puzzling string solution with non-integrable motion even in the zero backreaction limit. The section is supported by two appendices. In appendix A we present more details of the $\CPthree$ geometry, while in Appendix B we show how it is possible that the non-integrable $T^{11}$ space can arise as a foliation of $\CPthree$ without conflicting with integrability.

\section{String equations of motion for a generic metric} \label{section:gen}

In this section we generate the string equations of motion in a generic framework and study their simplifications
for a particular string parametrisation. The metric may be written as
\bea\nn
ds^2=g_{tt} \diff t^2 +g_{ii} \diff x^i \diff x^i+2 g_{ij} \diff x^i \diff x^j~,
\eea
where in the last term $i<j$ and the indices $i,j=1,\ldots,d$ with $d$ being the number of space dimensions and $t$ being the coordinate time. Let's assume that we have $n$ cyclic coordinates, such that the labels $i\ge n$ label the cyclic angles, on which the metric elements functions are not dependent. The Nambu-Goto action is (using $\frac{\partial}{\partial \tau}:=\dot{},\frac{\partial}{\partial\sigma}:={}^\prime$)
\be\nn
S=\int d\s d\t \prt{g_{tt} \dt^2+ g_{ii} \prt{ x^{i\prime}{~}^2-\dot{x}^{i}{~}^2 }+2 g_{ij} \prt{x^{i\prime} x^{j\prime}- \dot{x}^i\dot{x}^j}}:=\int d\s d\t \prt{g_{tt} \dt^2+ \cL_s}~.
\ee
We parametrise the string in such a way that all the non-cyclic coordinates $\a^i:=x^i$ with $i<d-n$ depend on the world-sheet time $\t$ and therefore the string is localised along those directions. We also take the string to extend along all the cyclic coordinates $\phi^k:=x^k$ with $k\ge n$ in a way that each of them is linear in the world-sheet space parameter $\s$ with proportionality constant $m_k$.
The equation of motion for the time $t$, for a static space reads
\be\la{gentau}
\dt= \frac{\k }{g_{tt}}~,
\ee
where  $g_{tt}$ is a function of $\t$. The equations of motion for the non-cyclic angles are
\be\label{eomnocycle}
\partial_{\a^i} \cL_s +2 \partial_\t \prt{g_{ij}\dot{\a}^j}=0~.
\ee
Notice the simplification in the absence of the $\s$ derivatives as  a result of the string parametrisation.
The simplest equations for the system are the ones for the $U\prt{1}$ angles reading
\be\label{eomcycle}
\partial_0 \prt{g_{\phi_i\phi_j}\dot{\phi_j}}- \partial_1 \prt{g_{\phi_i\phi_j}\phi_j'}=0~.
\ee
For the parametrisation we have considered these are satisfied trivially since the metric elements depend on the non-cyclic coordinates.

The Virasoro constraints become
\bea \label{vcg1}
g_{ij} \dot{x}^i x'^{j}= 0~,\\ \label{vcg2}
g_{tt} \dt^2+\cL_{s+}=0~,
\eea
where $\cL_{s+}$ is derived from the Lagrangian density expression by flipping the negative signs in front of the kinetic terms
to positive ones.

So far the non-trivial equations are \eq{gentau}, \eq{eomnocycle} and the Virasoro constraints \eq{vcg1} and \eq{vcg2}. A further simplification happens when the metric has no non-diagonal terms between the non-cyclic $\a_i$ and the cyclic angles $\phi_i$. Then the Virasoro constraint \eq{vcg1} is satisfied trivially and the equations \eq{eomnocycle} are summed only in the non-cyclic directions.

Below we will apply the general formalism developed in this section to the gravity dual theories under examination.

\section{Backreacted flavors in four dimensions}

\subsection{Dual geometry to $\Ncal=2$ SYM with unquenched flavor}

In this section we examine the $\Ncal=2$ super Yang-Mills theory with $N_f$ hypermultiplets in the fundamental representation and one in the adjoint representation of the $SU(N_c)$ gauge group. The gravity dual of this background is the near horizon limit of $N_c$ coincident branes with $N_f$ number of D7-branes sharing the four spacetime directions with the D3-branes and extending along four of the six transverse directions. In the quenched approximation, where the D7-branes do not backreact on the geometry, the theory was discussed in \cite{Karch:2002sh}. As the geometry in this limit is the same as the case of no flavors, integrability for closed strings is unaffected. Integrability
for the open string sector was established in \cite{Mann:2006rh}.  Here we are interested in going beyond the quenched approximation and study integrability of string motion
for the geometry dual to the unquenched theory, where the effects of backreaction of a large number $N_f$ of flavor branes are included.\footnote{One way to
obtain an exact background with flavors is through an orientifolding procedure, which leads to an $\AdS_5\times \Srm^5/\mathbb{Z}_2$ geometry. Integrability of
open strings in this setup was studied in \cite{Stefanski:2003qr,Chen:2004yf}. In this case the number of flavors is low, far from the $N_f\sim N_c$ regime that we are interested in.} Building on \cite{Aharony:1998xz, Grana:2001xn}, this geometry was constructed in \cite{Burrington:2004id,Kirsch:2005uy}.

The background has the metric
\be\la{metricd3d7}
\diff s^2=h^{-1/2}(r,\r) \diff x_\m \diff x^\m+ h^{1/2}(r,\r)\prt{\diff r^2+r^2 \diff \Omega_3^2+e^{\Psi(\r)}\prt{\diff\r^2+\r^2 ~ \diff\psi^2}}~,
\ee
where an $SO(4)\times SO(2)$ symmetry is present. The warp factor $h$ has been found in terms of a convergent series\footnote{In \cite{Aharony:1998xz} it was found to a first order approximation in the transverse 2-plane around a fixed point above the D7-branes.} in \cite{Kirsch:2005uy} following the methodology developed for the Schr\"{o}dinger equation of electrons in a logarithmic potential \cite{Gesztesy:1977vd}. Around the vicinity of the D7-branes and in the near horizon limit the warp factor becomes
\be\la{warp}
h(r,\rho)=\frac{Q_{D3}}{(r^2+\rho^2 e^{\Psi(\r)})^2}~,
\ee
with
\be
Q_{D3}=4\pi g_s N_c l_s^4 ,\quad \Psi(\r)=\log\prt{b_f \log\frac{\rho_L}{\rho}}~,\quad \rho_L:=e^{\frac{2\pi N_c}{\l N_f}}~,\quad b_f:=\frac{N_f}{2\pi}~.
\ee
The constant $\r_L$ is chosen such that in the absence of the flavors $e^\phi=g_s$. The warp factor expression is only valid for values away from $\r_L$, as $\Psi(\r)$ diverges around this value.

In the following sections we will parametrise the $\Srm^3$ of the metric \eq{metricd3d7} as
\be\la{param1}
\diff\Omega_3^2=\diff\th^2+\sin^2\th\prt{\diff\a^2+\sin^2\a \diff\chi^2}~,
\ee
while the four dimensional spacetime is parametrised as
\be\la{param2}
 \diff x_\m \diff x^\m= -\diff t^2+\diff R^2+R^2 \diff \phi^2+\diff z^2~,
\ee
which is a convenient choice to describe circular strings with radius $R$.\footnote{In the following, as in \cite{Kirsch:2005uy}
  and much of the literature on e.g. pulsating string solutions, we will allow $R$ and $r$ to take negative values, signifying a
string of opposite orientation.}

\subsection{String solutions and non-integrability}

In order to study (non)-integrability of the theory, we examine classical string solutions on the above background.
We are interested in string solutions that extend along the cyclic angles $(\phi,\chi,\psi)$ linearly with the $\s$ string
worldsheet parameter, where the linearity constants are set to be $(m_\phi,~m_\chi,~m_\psi)$. In other words, the string is
wrapping these angles an amount of times given by the corresponding $m$. All the other angles, as well as the coordinate
time itself, are only allowed to be functions of the time world-sheet parameter $\t$.

The equations of motion are derived by the Nambu-Goto action in the geometry of \eq{metricd3d7} with the parametrisations
\eq{param1} and \eq{param2}. Our parametrisation ensures that all the cyclic equations of motion are trivially
satisfied, while the time equation of motion gives
\be\la{timed7}
\dot{t}(\t)=\frac{\k}{2}\frac{\sqrt{Q_{D3}}}{|r(t)^2-b_f \log(\frac{\r(\tau)}{\r_L})\r(\tau)^2|}~,
\ee
where $\k$ is the integration constant.
By placing the string on the equators $(\th=\pi/2,~ \a=\pi/2)$ of the $\Srm^3$ internal sphere the two relevant equations are
satisfied. Moreover, the $z$ equation of motion is satisfied by switching off this direction for the string. We are now
left with the three equations of motion corresponding to $(R(\t),r(\t),\r(\t))$ functions and one non-trivial Virasoro constraint.

It is important to observe that the harmonic function $h$ contains the term $\log(\r(\t)) \r(\t)^2$ where its
derivative with respect to $\t$ or $\r(\t)$ gives the same factor $1+2 \log(\r(\t))$. Due to this fact the $\rho$
equation of motion is satisfied for $\rho= \r_L/\sqrt{e}$. Note that this value is within the regime where
the background approximation is valid. The remaining equations of motion are those for $R(\t)$ and $r(\t)$, which read
\bea\label{eaf1}
&&m_\phi^2\prt{b_f\r_L^2+2 e r(\t)^2}R(\t)+4 e r(\t)\dot{r}(\t) \dot{R}(\t)+\prt{b_f \r_L^2+2 e r(\t)^2}\ddot{R}(\t)=0~,\\\nn
&&\frac{2}{\sqrt{Q_{D3}}(b_f \r_L^2+2 e r(\t)^2)^2}\bigg( r(t)\Big(-m_\phi^2\prt{b_f \r_L^2+2 e r(\t)^2}^2 R(t)^2+e Q_{D3} (-\k^2 e+ 4 e m_\chi^2 +2 b_f m_{\psi}^2 \rho_L^2\\&&+4 e r'(t)^2) +\prt{b_f \rho_L^2 +2 e r(t)^2}^2 R'(t)^2\Big)-2 e Q_{D3} \prt{ b_f \r_L^2+2 e r(t)^2}r''(t)\bigg)=0~,\label{eaf2}
\eea
while the non-trivial Virasoro equation reads
\bea\nn \label{virf1}
&&\frac{1}{2 e \sqrt{Q_{D3}}(b_f \r_L^2+2 e r(\t)^2)} \bigg(m_\phi^2 (b_f \r_L^2+2 e r(\t)^2)^2 R(t)^2 + e Q_{D3} (-\k^2 e+ 4 e m_\chi^2\\
&&\hspace{4cm} +2 b_f m_{\psi}^2 \rho_L^2  +4 e r'(t)^2)+(b_f \r_L^2+2 e r(\t)^2)^2 R'(t)^2\bigg)=0~.
\eea
A string solution of the above system and its NVE does exist. To simplify the system significantly without losing generality, we localise the string on the internal sphere by setting $m_\chi=0$. This allows an analytic $r(\t)$ solution if the
 string is taken to move in the $R(\t)=0$ plane
\be
r(\t)=\frac{\sqrt{\k^2 e-2 b_f m_\psi^2 \r_L^2}}{2 \sqrt{e}} \t+c_1~,
\ee
where $c_1$ is an integration constant. By applying variations along the $R(\t)=0+\eta(\t)$ direction, the corresponding equation of motion gives an NVE with rational coefficients
\be\label{nve01}
\ddot{\eta}(\t)+\frac{2 \prt{\k^2 e -2 b_f m_\psi^2 \rho_L^2}\t}{(\k^2 e -2 b_f \rho_L^2 m_\psi^2 )\t^2+2 b_f \r_L^2}\dot{\eta}(\tau)+m_\phi^2 \eta(\t)=0~,
\ee
where we set $c_1=0$. The above equation does not have Liouvillian solutions and the Kovacic algorithm fails. Therefore this means that the flavor background in the unquenched limit is non-integrable and the quark loops destroy the integrability.

In fact one may even further simplify the string solution and the corresponding NVE and still show non-integrability.
The string may be localised consistently in the transverse $SO(2)$ plane, parametrised by $(\r,\psi)$, by setting
$m_\psi=0$. Choosing the appropriate initial conditions we get an invariant plane solution $R(\t)=0$ with
$r(\t)=\k~ \t$, and the variation of  $R(\t)=0+\eta(\t)$ leads to
\be\la{nvesimple}
\ddot{\eta}(\t)+\frac{2 \k^2 e \t}{2 b_f \r_L^2+e \k^2 \t^2}\dot{\eta}(\t)+m_\phi^2 \eta(\t)=0~,
\ee
which can be also obtained directly from \eq{nve01}.
The minimal string NVE \eq{nvesimple} does not have a Liouvillian solution proving non-integrability in the strong coupling limit flavor backreacted backgrounds.

Integrability is recovered for particle motion $(m_\phi=0)$ where the NVE admits Liouvillian solutions
\be\label{etaparticle}
\eta(\t)=c_3+c_4 \frac{\arctan(\frac{\k \sqrt{e}\t}{\sqrt{2 b_f} \r_L})}{\k  \sqrt{2 e b_f}\r_L}~,
\ee
where $c_3$ and $c_4$ are the integration constants.

A comment is in order for the quenched limit of our solution. The expression
\be\label{parameter1}
b_f \rho_L^2=\frac{N_f}{2\pi}  e^{\frac{2\pi N_c}{\l N_f}}~,
\ee
is infinite both for small and for large number of flavors, while for $N_f=2\pi N_c/\l$ the function has a minimum at $ e N_c/\l$,  depending on the number of colors (and flavors) as well as the choice of the t'Hooft coupling. Moreover the string position in this limit goes deeply in the IR as $\rho\to \infty$ when considering the t'Hooft limit. As a result the NVE \eq{nve01} for $N_f=0$ does not have a smooth limit to the unquenched approximation. This limit may be recovered only together with the point-like string motion giving the integrable motion \eq{etaparticle}.

To summarise this section, we showed the existence of non-integrable string solutions in the flavor-backreacted D3/D7 background of \cite{Kirsch:2005uy}. Integrability is recovered for particle motion, a situation that has been observed in several other non-integrable theories. In the next section we study the existence of chaos in the flavor backreacted system.

\begin{figure}
\begin{minipage}[ht]{0.5\textwidth}
\begin{flushleft}
\centerline{\includegraphics[width=60mm]{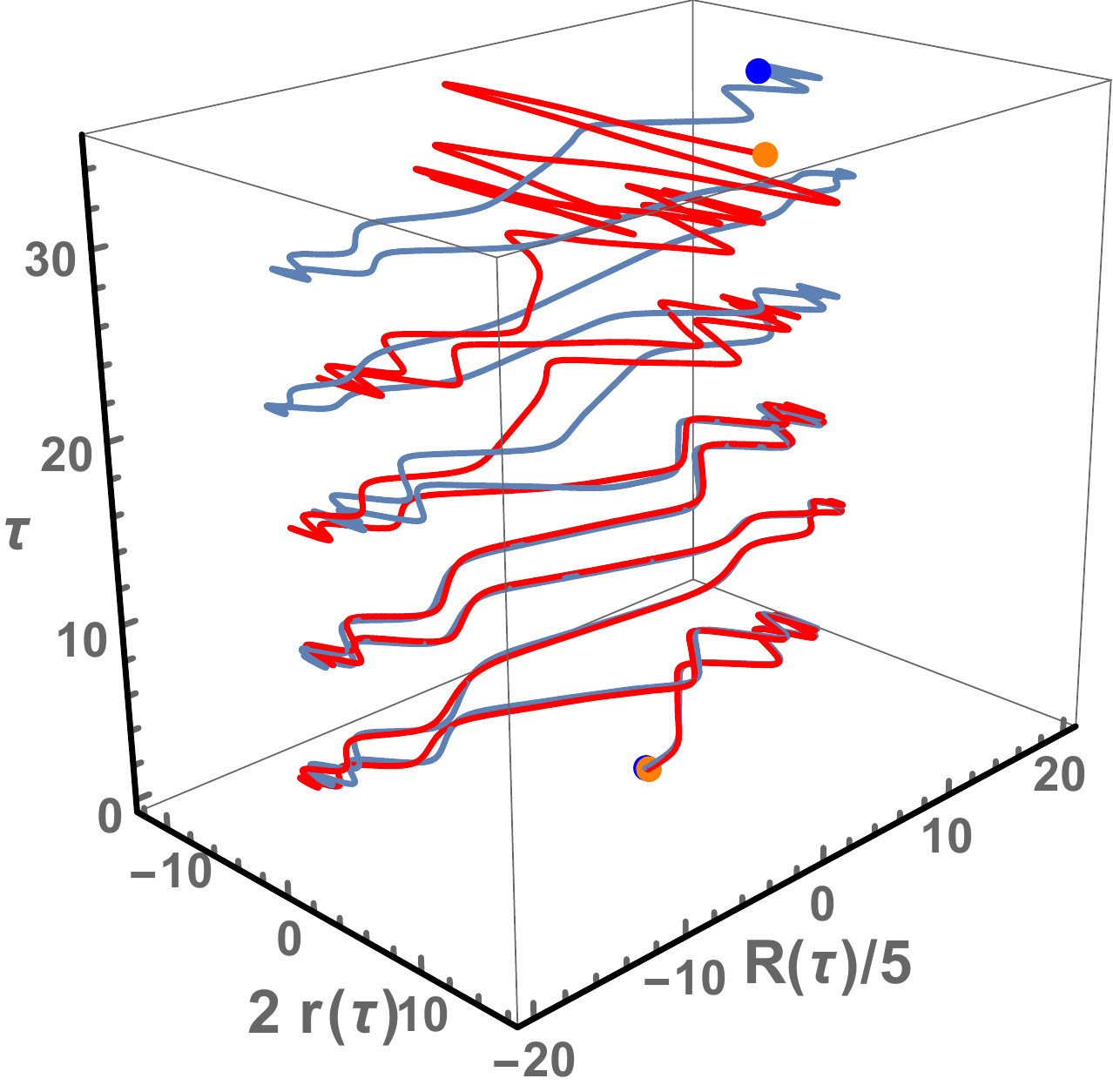}}
\caption{\small{The evolution of string motion along the holographic direction and its radius. The two strings have close initial conditions while the final endpoints begin to differ significantly. We simulate the motion until the initial moments of the chaos appearance. The axes have been rescaled for better optical results. }}
\label{figure:a1}
\end{flushleft}
\end{minipage}
\hspace{0.3cm}
\begin{minipage}[ht]{0.5\textwidth}
\begin{flushleft}
\centerline{\includegraphics[width=75mm ]{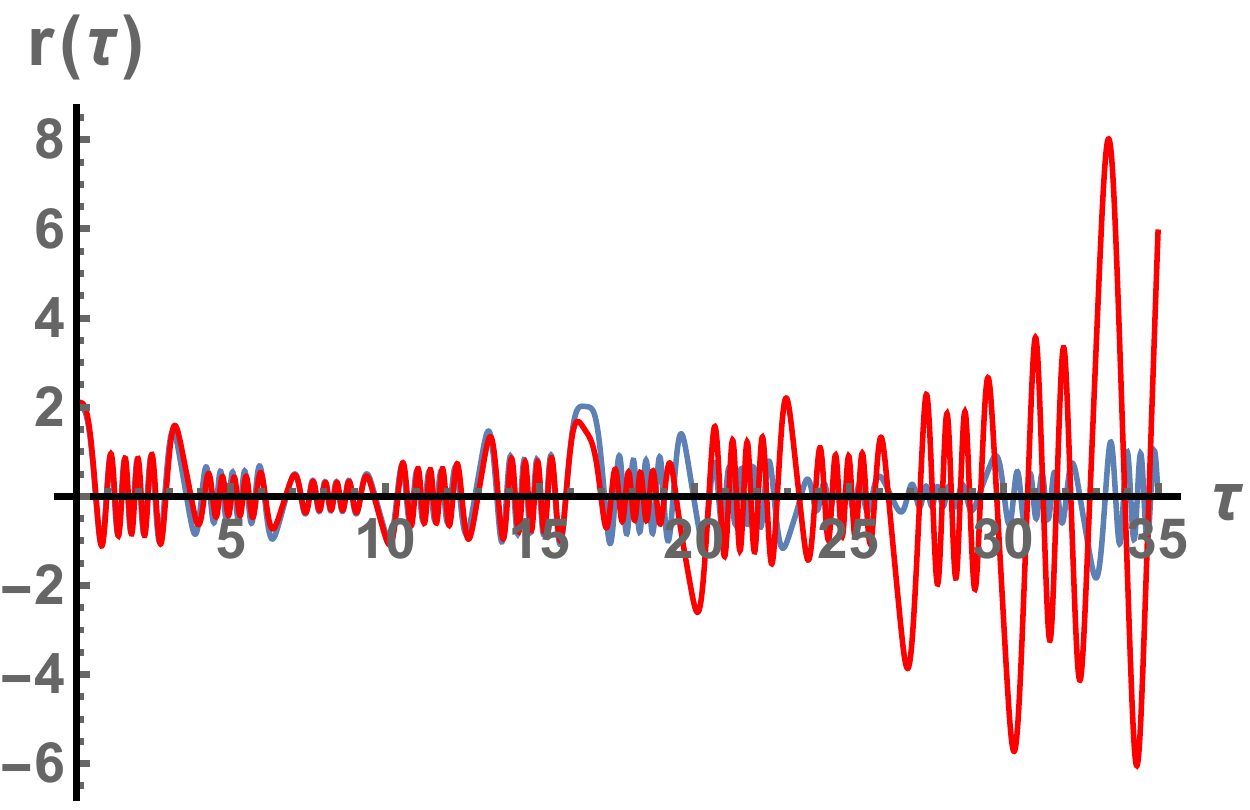}}
\caption{\small{The evolution of the strings along the direction $r$ for the same initial conditions as in Figure \ref{figure:a1}. Notice that the negative values of $r(\t)$ and $R(\t)$ correspond to radii of $|r(\t)|$ and $|R(\t)|$ with changed string orientation comparing to the positive values. Both plots have $N_f/N_c\sim 0.3,~\lambda=12$, where the values have chosen for best optical results. Higher values of $N_f/N_c$ give a similar pattern and make the chaotic oscillations more severe and observed at earlier times.} }
\label{figure:a2}\vspace{.0cm}
\end{flushleft}
\end{minipage}
\end{figure}

\subsection{Chaotic string motion}

Having shown analytically the non-integrability of string motion on the backreacted D3/D7 background, we examine the
presence of chaos in this background. To do this, we consider the same string ansatz as above but without choosing any specific solution for $R(\t)$. So the fields that will have a nontrivial motion are $r(\t)$ and $R(\t)$, corresponding to a time
evolution of a boundary operator. In terms of the Hamiltonian the system is reduced to a particle motion with four
dynamical variables $(r,R,\dr,\dR)$ and certain constant parameters. The conserved quantity of the system is the total energy. We look for chaos by varying the initial conditions and the theory parameters
resulting in modifying the energy density, and each time we integrate the equations of motion numerically.

Let us fix the value of the 't Hooft coupling $\lambda$ and examine the appearance of chaos as a function of $N_c/N_f$. We solve the system of differential equations \eq{eaf1} and \eq{eaf2}, with the energy constrained by \eq{virf1}. The 3-dimensional string motion with two neighbouring initial conditions is presented in Figure \ref{figure:a1}, where the chaotic nature starts to become apparent as the time evolves.  In Figure  \ref{figure:a2} we project the string motion to one dimension $r(\t)$ where the chaotic nature also begins to be apparent.

\begin{figure*}[!ht]
\centerline{\includegraphics[width=140mm]{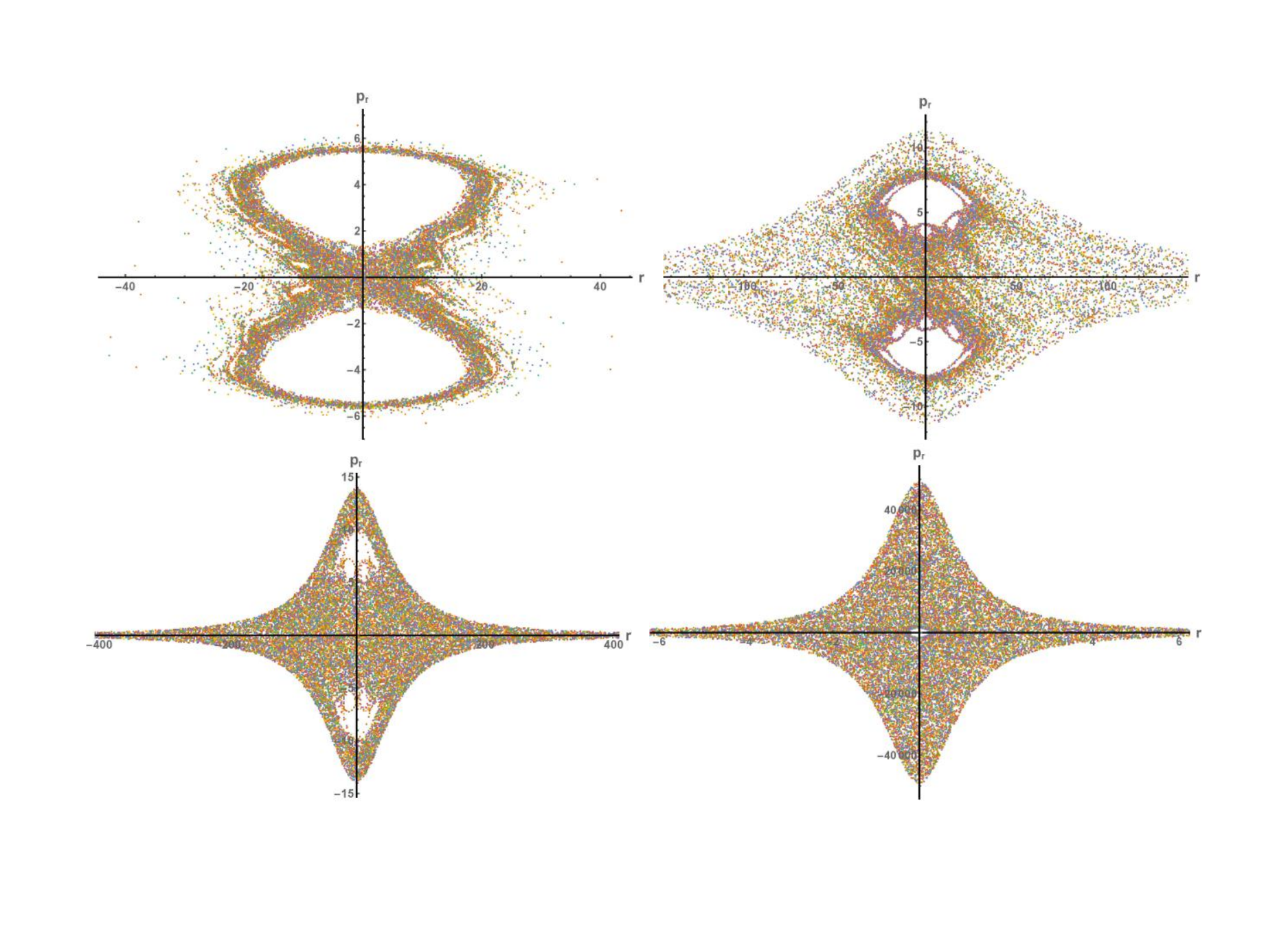}}
\caption{\small{The Poincar\'e sections for the flavor backreacted theory. The section is chosen such that $r(\t)=0$. The constant $\k=140$, the winding number $m=1$ and $\lambda/N_c$ is around unity. The bottom right plot is the Poincar\'e section for $N_f/N_c\sim 1$ where the chaos is obvious. The other three plots from top left to right correspond to $N_f/N_c=0.094,~0.096,~0.098$. Note that at this limit the range of the validity of the gravity solution is already challenged. However we allow ourselves to push towards this limit to illustrate the formation of chaos. }}
\label{figure:a3}
\end{figure*}

The chaotic motion can be quantified in terms of $N_c, ~N_f$ and $\l$, parameters that our system and the equations of motion depends on through $b_f \rho_L^2$ and is equal to \eq{parameter1}. The integrability of our equation is recovered in the limit $N_c/(\l N_f)=1/(2\pi)$  with $N_f\sim N_c/\l\to 0$, which is beyond the regime of validity of our background.
 A way to observe how chaos depends on the parameters of the system is by the construction of the Poincar\'e sections which are presented in Figure \ref{figure:a3} for an increasing number of flavors.
The choice of the constant $\k$ plays a crucial role in the energy density and has to be chosen in such a
way as to allow consistent initial conditions. We are already in the chaotic region and therefore we cannot
clearly observe the destruction of tori. However by slightly extending our analysis outside the
appropriate limit for $N_f/N_c$ and considering a low number of flavors  we get a picture of how chaos starts
to form.

To quantify the sensitivity of our system to the initial conditions we compute the leading Lyapunov exponent. We obtain the exponent by a long-time numerical calculation of two initially neighboring trajectories, where each time we measure the exponent, we rescale the distance in order to keep the nearby trajectory separation within the linearised flow range. Our main finding  is that when  $N_f\sim N_c$
the Lyapunov coefficient depends weakly on the increase of the number of flavors with a slow increasing trend. It almost converges to a value depending on the energy of the system. In the computation we vary $N_f$ by pushing again
the limits of our background validity. This is tempting since we find that for low values of flavor numbers we see
lower Lyapunov coefficient, which nevertheless quickly converge to the particular value. The Lyapunov exponent in terms of
the ratio $N_f/N_c$ can be well approximated with shifted sigmoid functions like the $\arctan$.

\begin{figure}
\centerline{\includegraphics[width=100mm]{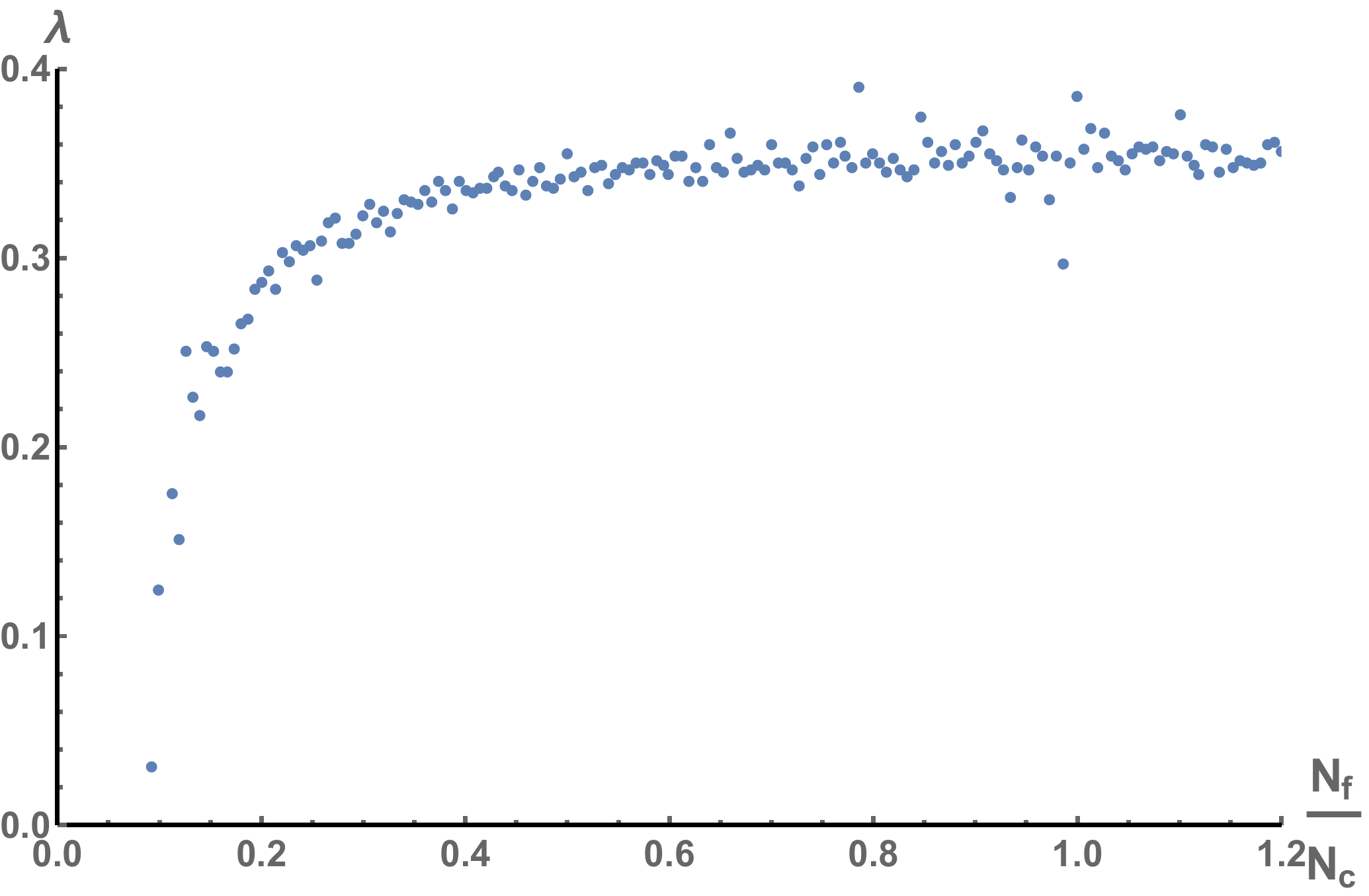}}
\caption{\small{The leading Lyapunov exponent in terms of number of flavors over number of colors. The main result is that in the region $N_f\sim N_c$ the exponent quantifying the chaotic motion is weakly depending on the number of flavors. It is tempting to extend our computation   for lower values of flavor pushing the limits of the validity of our background. Starting from small values
of $N_f/N_c$ and increasing, we see that the exponent quickly converges to the constant value already at $N_f\sim 0.4 N_c$. }}
\label{figure:a4}
\end{figure}

\section{Backreacted flavors in three dimensions}

In this section we consider the integrability of string motion on  backgrounds dual to three-dimensional
gauge theories with unquenched flavor. These backgrounds arise by adding branes to the $\Ncal=6$ ABJM
Chern-Simons-matter theory, which is known to be integrable
(see \cite{Arutyunov:2008if,Stefanski:2008ik} for
the Lax-pair construction for the string sigma model),
and allowing them to backreact. In the following we review the geometry of the ABJM background and discuss how to
obtain the deformation corresponding to backreacted flavor, before turning to the study of string solutions
relevant for the study of integrability.

\subsection{The ABJM background} \label{Section:ABJM}

The ABJM background \cite{Aharony:2008ug} is given by
\be \label{metric01}
\diff s^2=L^2 \diff s^2_{\AdS_4}+4 L^2 \diff s^2_{\CPthree}~,
\ee
where $L^4=2 \pi^2 \frac{N_c}{k}$, and the dilaton,  RR 2- and 4-form field strengths are given by
\be
e^\phi=\frac{2 L}{k}=2 \sqrt{\pi}\prt{\frac{2 N}{k^5}}^\frac{1}{4}~,
\quad F_2=2 k J~,\quad F_4=\frac{3}{2} k L^2 \Omega_{\AdS_4}=\frac{3 \pi}{\sqrt{2}}\sqrt{k N} \Omega_{\AdS_4}~,
\ee
where $J$ is the K\"{a}hler form of $\CPthree$ and $\l=N_c/k$ is the t'Hooft coupling. This is dual to a Chern-Simons-matter theory,
of rank $N_c$ and at level $k$, in the limit $N_c^\frac{1}{5}\ll k\ll N_c$, corresponding to
large $\AdS$ radius and small string coupling.

We will mostly be concerned with the internal $\CPthree$ space, so we will provide some details
on its geometry. The form of $\CPthree$ which is convenient for introducing the backreaction is as an $\Srm^2$-bundle over $\Srm^4$, using the self-dual $\SU\prt{2}$ instanton on the four-sphere
\be\label{metrica1}
\diff s^2_{\CPthree}=\frac{1}{4}\prt{\diff s^2_{\Srm^4}+\prt{\diff x^i+\e^{ijk}A^j x^k}^2}~.
\ee
Here the $x^i$ are the Cartesian coordinates of the $\Srm^2$ satisfying $x_i^2=1$ and $A^i$ is the non-abelian one-form
connection of the $\SU\prt{2}$ instanton. The metric of the four-sphere reads\footnote{In the parametrisation of
\cite{Conde:2011sw} a coordinate $\xi$ is used instead of $\alpha$, with the relation between these coordinates
being $\sin\a=\frac{2 \xi}{1+\xi^2}$, or $\xi=\tan\frac{\alpha}{2}$.
}
\be
\diff s^2_{\mathbb{S}^4} =
\Big[\diff \a^2+\frac{1}{4}\sin^2\a \,\sum_{i=1}^3 ( \omega^i)^2\,\Big]\,\,,
\ee
where the range of $\a$ is $0\leq\a \le \pi$
and the $\omega_i$ are $\SU\prt{2}$ left-invariant forms satisfying
$\diff \omega^i=\frac{1}{2} \e_{ijk} \o^j\wedge \o^k$. We will write them explicitly in terms of angular coordinates as
\bea \label{leftinv1}
\omega^1 & = & \cos\psi_1\,\diff\theta_1+\sin\psi_1\,\sin\theta_1\,\diff\varphi_1~,\\ \label{leftinv2}
\omega^2 & = & \sin\psi_1\,\diff\theta_1-\cos\psi_1\,\sin\theta_1\,\diff\varphi_1~, \\ \label{leftinv3}
\omega^3 & = & \diff\psi_1+\cos\theta_1 \,\diff\varphi_1\,\, .
\label{omegas}
\eea
Then the one-forms $A^i$ are expressed as
\be
A^i=-\sin^2\frac{\a}{2}\omega^i~,
\ee
while the $x^i$ are parametrised with the spherical angles $\th$ and $\varphi$ giving
$\prt{\diff x^i+\e^{ijk}A^j x^k}^2= \prt{E^1}^2+\prt{E^2}^2$.
The $E^1$ and $E^2$ are two one-forms that can be written as
\bea
E^1 & = & \diff\theta+\sin^2\frac{\a}{2}\prt{\sin\varphi~\omega^1 -\cos\varphi~\omega^2}~, \\
E^2 & = & \sin\theta\prt{\diff\varphi-\sin^2\frac{\a}{2}\omega^3}
+\sin^2\frac{\a}{2}\cos\theta\prt{ \cos\varphi\omega^1+\sin\varphi\omega^2} ~ .
\eea
The RR two-form can be written as
\be
F_2=\frac{k}{2}\prt{E^1\wedge E^2-\prt{\cS^\a\wedge \cS^3+\cS^1\wedge \cS^2}}~,
\ee
where
\bea\nn
&&\cS^\a= \diff\a~,\qquad \cS^i=\frac{\sin{\a}}{2}S^i~,\\\nn
&&S^1=\sin\varphi \omega^1-\cos\varphi \omega^2~,\quad
S^2=\sin\theta \omega^3-\cos\theta\prt{\cos\varphi\,\omega^1+
\sin\varphi\,\omega^2}~, \\
&&
S^3=-\cos\theta \omega^3-\sin\theta\prt{\cos\varphi\,\omega^1+
\sin\varphi\,\omega^2}~.
\la{si}
\eea
The addition of flavor D6 branes to the ABJM background, in the probe approximation, was
considered in \cite{Hohenegger:2009as,Gaiotto:2009tk,Hikida:2009tp,Jensen:2010vx,Ammon:2009wc,Zafrir:2012yg}. The
D6-branes are wrapped around an $\RPthree$ subspace of $\CPthree$, with the configuration preserving $\Ncal=3$ supersymmetry. The integrability of the corresponding gauge theory (taking into account open spin chains corresponding
to operators with fundamental fields) was recently discussed in \cite{Bai:2017jpe}.
In this work we are interested in the geometry arising when the backreaction of the flavor branes is taken into account,
corresponding to unquenched flavors in the ABJM gauge theory. We will discuss the resulting geometry in the next section.

\subsection{The flavor-backreacted background}

The deformed $\AdS_4\times \CPthree$
metric derived in \cite{Conde:2011sw} is
\be\label{metric1}
 \diff s_{10}^2 =
 L^2\prt{-\cosh^2\rho ~\diff t^2+\diff \rho^2+\sinh^2\rho~ \diff \Omega_2^2} +  \frac{L^2}{b^2}\left(q \diff s^2_{\mathbb{S}^4}+(E^1)^2+(E^2)^2\right) \ ,
\ee
where the $\AdS_4$ metric is written in global coordinates and $q$ and $b$ are constants measuring the degree of the deformation from the original metric, by changing the size of the $\Srm^4$ compared with $\Srm^2$ of the $\CPthree$ manifold. The special case of $q=1$ corresponds to the undeformed $\cN=6$ supersymmetric ABJM background, while for all other values of $q$ the space is squashed. The other special case $q=5$ corresponds to an $\cN=1$ supergravity background, dual to superconformal Chern-Simons matter theory with $\SO\prt{5}\times \Urm\prt{1}$ global symmetry \cite{Ooguri:2008dk}. The parameter $b$ can be thought as the scaling of the energy of a localised observable in the AdS space.

The flavored $\AdS$ solutions for constant parameters are given by the solution of the BPS conditions as
\bea \label{qvalues}
 q = 3+\frac{3}{2}\hat\epsilon \pm 2\sqrt{1+\hat\epsilon+\frac{9}{16}\hat\epsilon^2}~,\qquad
 b = \frac{2q}{q+1}\ ,\label{constants}
\eea
where $\hat\epsilon$ is the usual Veneziano parameter reading
\be
 \hat \epsilon =\frac{3}{4}\frac{N_f}{N_c}\lambda \ .
 \label{epsilon}
\ee
The branch with the minus sign extends from $q=1$ to $5/3$ and is the flavored backreacted ABJM, since $q=1$ is included. The branch with positive sign extends from $q=5$ upwards and corresponds to the flavored version of the deformed $\cN=1$ CS matter theory, since the value $q=5$ is included.

The mentioned backgrounds appear also as the IR fixed points of D2-D6 brane system flows. By computing the backreaction
of the $N_f$ flavor D6-branes smeared over a six-dimensional nearly K\"{a}hler manifold, to the $N_c$ color branes, it has
been found that the solutions flow to an $\AdS_4$ fixed point dual to Chern-Simons-matter theory and which is the special
case of the metric \eq{metric1}  for  a specific  $q$-value and a squashing parameter appearing in  the two-form \cite{Faedo:2015ula}.

It is important to note that the flavor-backreacted geometry contains an $\AdS_4$ factor, as the addition of
fundamental flavors to ABJM does not generically break conformal invariance (see
\cite{Bianchi:2009ja,Bianchi:2009rf} for perturbative gauge theory considerations). This is unlike the
four-dimensional case discussed above and of course results in a much simpler geometric
background.

Since the construction of the above background involved smearing D6-branes with different orientations,
the resulting supersymmetry is $\Ncal=1$ in three dimensions,
instead of $\Ncal=3$, which would only be expected for the supergravity solution with
localised flavor branes \cite{Conde:2011sw}.

\subsection{String Solutions on the backreacted metric}

In this section we explore the integrability of string motion on the flavor backreacted ABJM background above.
We will consider three ans\"atze, which are distinguished by whether we set $\theta=0$ ($\theta$ being
the angle of the $\Srm^2$), $\theta=\pi/2$,  or $\theta_1=0$ ($\theta_1$ being the angle of the $\Srm^4$).

\subsubsection{Static String on $\Srm^2$}

\begin{figure}[h]
\begin{center}
\resizebox{2.0in}{!}{\includegraphics{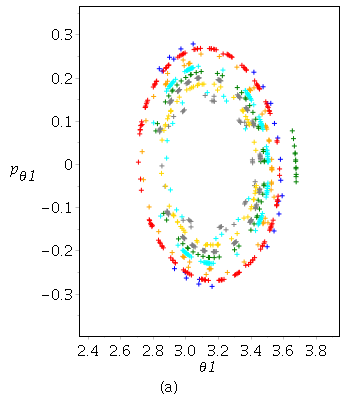}} \,\,
\resizebox{2.0in}{!}{\includegraphics{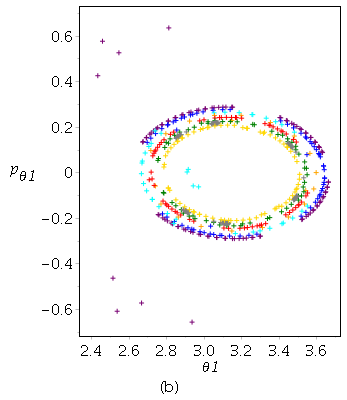}}\,\,
\resizebox{2.0in}{!}{\includegraphics{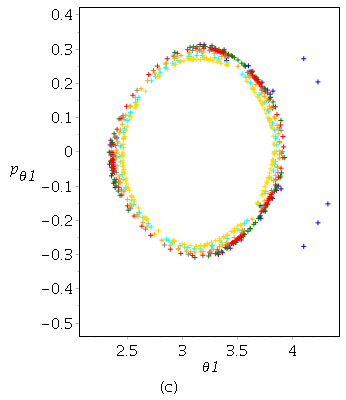}}\vspace{-.2cm}
\caption{ Sample Poincar\'e plots for the ansatz (\ref{Theta1Ansatz1}), here plotted for $m=1,\kappa=0.6$, and varying values of the deformation parameter
$q$. (a) $q=1$, (b) $q=1.1$ and (c) $q=1.65$. The total energy is restricted to $H=0$ due to the Virasoro constraint.} \label{Theta0ansatz}
\end{center}
\end{figure}

We will now consider an ansatz for string motion which will turn out to be suitable for the application of analytic non-integrability techniques. We choose:
\be \label{Theta1Ansatz1}
t=\kappa\tau~,\quad \a=\a\prt{\t}~,\quad \theta=0~, \quad \theta_1=\theta_1\prt{\t}~,\quad \varphi_1= m\s~,\quad \varphi=\frac{\pi}{2}~,\quad\psi_1=\frac{\pi}{2}\;.
\ee
As this parametrisation is contained within the general analysis of section \ref{section:gen}, it is immediate to analyse
the equations of motion. Doing this we find that the only non-trivial equations of motion are those for $\theta_1$ and $\alpha$
\be
\begin{split}
\partial_\a\cL_s+2\partial_0\prt{g_{\a\a}\dot{\a}}=0~,\\
\partial_{\theta_1} \cL_s+2\partial_0\prt{g_{\theta_1\theta_1}\dot{\theta_1}}=0~,\\
g_{tt} \dot{t}^2+\cL_{s+}=0~.
\end{split}
\ee
Writing these out more explicitly, we obtain
\be
q \ddot \alpha-\frac14{\dot\theta_1}^2\sin\alpha(1+\cos\alpha(q-1))
+\frac14m^2\sin\alpha\left(q \cos\alpha+(1-\cos\alpha)\sin^2\theta_1\right)=0
\ee
and
\be
\ddot\theta_1(1+q\sin^2\alpha+\cos^2\alpha-2\cos\alpha)+2\dot\theta_1\dot\alpha\sin\alpha (1+(q-1)\cos\alpha)
+ m^2\sin\theta_1\cos\theta_1(\cos\alpha-1)^2=0~.
\ee
The two equations of motion may be thought as coming from an effective particle lagrangian
\be
\begin{split}
\Lcal_{\text{eff}}&=b^2\kappa^2+\frac{q}{4}\alphadot^2+\frac{1}{16}\left(4 \sin^4\frac{\a}{2}+q \sin^2\a \right)\thetadot_1^2 -\frac{m^2}{16}\prt{4 \sin^2\th_1 \sin^4\frac{\alpha}{2}+q\sin^2\alpha}~.
\end{split}
\ee
Converting to the Hamiltonian using $p_\alpha=\partial \Lcal_{\text{eff}}/\partial \alphadot$, $p_{\theta_1}=\partial \Lcal_{\text{eff}}/\partial \thetadot_1$,
we obtain
\be
\begin{split}
\Hcal&=\frac{p_\alpha^2}{q}+\frac{4 p_{\theta_1}^2}{4 \sin^4\frac{\a}{2}+q \sin^2\a}-b^2\kappa^2+\frac{ m^2}{16}\prt{4 \sin^2\th_1 \sin^4\frac{\alpha}{2}+q\sin^2\alpha}~.
\end{split}
\ee
The Virasoro constraint sets the energy to zero, which must be imposed on the dynamics of this Hamiltonian. Using these formulas we can treat the system as one of particle motion and construct the Poincar\'e maps as  in Figure \ref{Theta0ansatz}, which presents different values of $q$. We find that, at least for the initial conditions we have considered, there is only weak evidence of chaotic behaviour in this parameter space. This is confirmed by a computation of the Lyapunov exponent,
which, although positive, turns out to be small and not significant enough to make a conclusive statement about chaos in this system.

In the following we will turn to the techniques of analytic non-integrability, for which we need to choose an
invariant plane. We will consider two cases.

\textbf{ Case I: The $\theta_1=0$ plane}

On this plane, the only non-trivial equation of motion is  for $\alpha(\tau)$:
\be
\ddot\alpha+\frac 14 m^2 \sin\alpha\cos\alpha=0
\ee
which, using the Virasoro constraint, gives
\be
\dot\alpha^2=\frac{1}{4q}\left(16b^2\kappa^2-m^2q\sin^2\alpha\right)~.
\ee
To obtain the NVE, we expand $\theta_1(\tau)=0+\eta(\tau)$ and arrive at
\be
\frac12\ddot \eta \sin^2\frac{\alpha}{2}(1+q+(q-1) \cos\a)
+\frac12\dot\eta\dot\alpha\sin\alpha (1+(q-1)\cos\alpha)
+  m^2\sin^4\frac{\a}{2}\eta=0~.
\ee
We need to bring this equation into a form suitable for applying the Kovacic algorithm. In particular it needs to be
an ODE with rational coefficients. We can achieve this by making the substitution $z=\cos\alpha$, after which we find the
NVE\footnote{Primes denote $\frac{\partial}{\partial z}$.}
\be
\begin{split}\nn
&\eta''(z-1)^2(z+1)\prt{(q-1)z+q+1}\prt{m^2 q( z^2-1)+16 b^2 \kappa^2}+2\eta'(z-1)\big( m^2 q(z^2-1)\\
  &\cdot(1+z+2 q z+2(q-1)z^2)+8b^2k^2(2+z+3 q z +3(q-1)z^2)\big)
+4\eta q m^2 (z-1)^2=0~.
\end{split}
\ee
Applying the Kovacic algorithm we find that this NVE does not admit integrable solutions for generic $q,m$.
However, specialising to $m=0$ (the point-like limit) or $q=1$ (the undeformed $\CPthree$ limit) we find
an integrable NVE. This is consistent with the integrability of geodesic motion on the backreacted backgrounds
as well as the integrability of the ABJM theory.

\textbf{Case II: The $\theta_1=\frac{\pi}2$ plane}

The other invariant plane in our problem is $\theta_1=\frac{\pi}{2}$. On this plane the $\alpha$ equations
of motion become
\be
\ddot \alpha=-\frac{m^2\sin\alpha}{4q}\left(1+(q-1)\cos\alpha\right)~.
\ee
Expanding $\theta_1=\frac{\pi}2+\eta(\tau)$ and changing variables to $z=\cos\alpha$, we find the NVE
\be
\begin{split}
&\eta''(z+1)(z-1)^2\prt{1+q +z(q-1)}(16 b^2 \kappa^2+m^2(z-1)(1+q +z(q-1))\\
&+\eta' (z-1)\cdot \bigg(16\kappa^2 b^2\prt{2+z+3 q z+3(q-1)z^2}+m^2(z-1)\prt{1+q+(q-1)z}\\
&\cdot \prt{3+z+4 q z+4 (q-1) z^2}\bigg)-4 m^2 q (z-1)^2 \eta=0~.
\end{split}
\ee
As above, we find that the Kovacic algorithm concludes for $m=0$ and for $q=1$, while it does not conclude with Liouvillian solutions in the generic case. We see that the backreacted background ($q>1$) of \cite{Conde:2011sw} is non-integrable, while for $q=1$ integrability appears to be recovered, for this string solution at least. In the next section we present the analysis of a string moving in a $\CPtwo$ subspace of $\CPthree$.

\subsubsection{The string on $\CPtwo$} \label{CP2section}

Let us now consider an ansatz corresponding to string motion purely on the $\CPtwo$ part of the
deformed $\CPthree$ spacetime. As discussed in appendix \ref{CP2restriction}, the restriction to the subspace defined by $\theta=\phi=\pi/2$, of the undeformed $\CPthree$ metric (in the $\Srm^2\rightarrow \CPthree\rightarrow \Srm^4$ fibration picture, where $\theta,\phi$ are the coordinates of the $\Srm^2$) leaves us with the Fubini-Study metric on $\CPtwo$. It is thus interesting to consider strings moving in this subspace both in the quenched and unquenched limit.

The string is parametrised as
\be \label{CP2ansatz}
t=\kappa\tau\;,\quad \theta=\phi =\frac{\pi}2,\quad \alpha =\alpha(\tau),\quad\theta_1 =\theta_1(\tau),\quad\phi_1 =m\sigma, \quad\psi_1 =\frac{\pi}{2}~,
\ee
giving non-trivial equations of motion for $\alpha$ and $\theta_1$ which are
\be
-\frac{q}{2}\alphadotdot+\frac{q}{8}\sin\alpha\cos\alpha\thetadot_1^2-\frac{q m^2}{8}\sin\alpha\cos\alpha
-\frac{m^2}{8}(1-\cos\alpha)\sin\alpha=0~,
\ee
while the $\theta_1$ EOM:
\be
-\frac{1}{8}\sin^2\alpha \thetadotdot_1-\frac{1}{4}\sin\alpha\cos\alpha \alphadot \thetadot_1=0~.
\ee
We note that $q$ does not enter into this equation.

The system can be reduced to a particle one with effective lagrangian reading
\be
\Lcal_{\text{eff}}=\frac{q}{4}\alphadot^2+\frac{q}{16}\sin^2\alpha\thetadot_1^2+b^2\kappa^2-\frac{q m^2}{16}\sin^2\alpha
-\frac{m^2}{16}(1-\cos\alpha)^2
\ee
and the corresponding Hamiltonian
\be
\Hcal=\frac{1}{q}p_\alpha^2+\frac{4}{q\sin^2\alpha}p^2_{\theta_1}-b^2\kappa^2+\frac{q m^2}{16}\sin^2\alpha
+\frac{m^2}{16}(1-\cos\alpha)^2~,
\ee
where we note that $\theta_1$ does not appear, only its conjugate momentum. Having these we can look at two nearby trajectories.
\begin{figure}[h]
\begin{center}
\resizebox{2.0in}{!}{\includegraphics{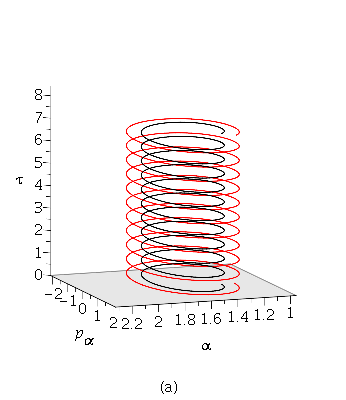}} \,
\resizebox{2.0in}{!}{\includegraphics{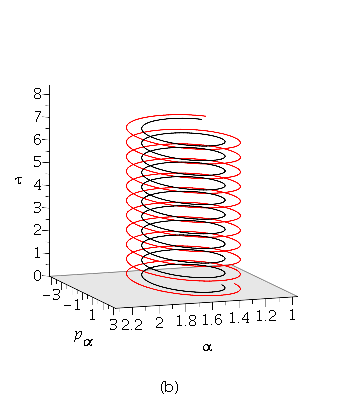}}\,
\resizebox{2.0in}{!}{\includegraphics{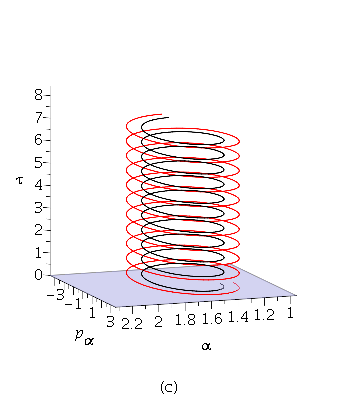}}\vspace{-.2cm}
\caption{ Phase space trajectories with nearby initial conditions for the $\CPtwo$ ansatz, here plotted for $m=1,\kappa=5$, and varying values of the deformation parameter
$q$. (a) $q=1$, (b) $q=3/2$ and (c) $q=5/3$. The total energy is restricted to $H=0$ due to the Virasoro constraint.} \label{CP2ansatzplots}
\end{center}
\end{figure}
From Fig. \ref{CP2ansatzplots} we see that motion on $\CPtwo$ is consistent with integrability, as expected. This is true for both $q=1$ and $q\neq 1$ in this simple string solution.

We now proceed to choose an invariant plane and check the non-integrability analytically. We take  $\theta_1=0$. The $\theta_1$ equation of
motion is automatically satisfied, while the $\alpha$ equation of motion becomes:
\be
\alphadotdot=-\frac{m^2}{4q}\sin\alpha((q-1)\cos\alpha+1)
\ee
and from the Virasoro constraints we find
\be
\alphadot^2=-\frac{1}{4q}\left(m^2(q\sin^2\alpha+\cos^2\alpha-2\cos\alpha+1)-16b^2\kappa^2\right)~.
\ee
Expanding $\theta_1(\tau)=0+\eta(\tau)$, keeping the first order in $\eta$, changing
variables from $\eta(\tau)$ to $\eta(z)$, where $z=\cos\alpha$, and substituting the above, we finally
find the NVE:
\be
\begin{split}
(z-1)(z+1)&\left(m^2(z-1)(qz+q-z+1)+16b\kappa^2\right)\ddot{\eta}(z)\\
&+\left((z-1)(4qz^2+4qz-4z^2+3z+1)m^2+48b^2\kappa^2z\right)\dot{\eta}(z)=0~.
\end{split}
\ee
Applying the Kovacic algorithm to this NVE, we find that it completes, consistent with string motion on
the $\theta=\phi=\pi/2$ subspace being integrable for any $q$, as the numerical results above also indicate.
For $q=1$ this agrees with our expectations
of integrability, but of course for $q\neq 1$ this could just be an artifact of our string ansatz being too simple
to show any non-integrability. To actually prove integrability on this subspace one would need to construct a
Lax pair for the metric (\ref{CP2qnot1}), similarly to \cite{Arutyunov:2008if,Stefanski:2008ik}. It would be
interesting to study string motion on this deformed $\CPtwo$ further in order to prove or disprove integrability.

In the next section we present the analysis of a different ansatz for string motion on the backreacted background,
which exhibits puzzling behavior in the  undeformed $\CPthree$ limit.

\subsubsection{The localised string on $\Srm^2$} \label{S2localised}

In this section we consider a generic parametrisation of string motion such that the "induced" $\SU(2)$ left-invariant forms  are $\o_2=\o_3=0~$, while $\o_3=\diff \varphi_1$.
The metric effectively becomes
\be \label{effectiveansatz1}
\begin{split}
\diff s^2=-\cosh^2\r ~\diff t^2+&\frac{1}{b^2}\Big(\diff \th^2 + q  \diff  \alpha^2+  \sin^2\th \diff \varphi^2
+\prt{ \frac{q}{4} \sin^2\a +\sin^4\frac{\a}{2}\sin^2\th}\diff \varphi_1^2\\
&\quad-2 \sin^2\frac{\a}{2}\sin^2\th \diff\varphi \diff\varphi_1\Big)~.
\end{split}
\ee
To restrict the string motion consistently to the above space we localise the string in AdS at $\r=0$ and set $\theta_1= \psi_1=0~$ and for further convenience $\vphi=0$.  By applying the formalism of section \ref{section:gen} to the full system of equations using the above parametrisation we may specify for the trivial part of the  solution
\be
 t=\k \t~,\quad \a=\a(\t)~,\quad \th=\th(\t)~,\quad \varphi_1 =m\s ~.
 \ee
To specify the remaining functions $\a$ and $\th$ we look at their equations of motion \eq{eomnocycle}
\bea\la{xi1a}
&&\ddot \a +\frac{m^2}{4 q} \prt{q \cos \a+2 \sin^2\frac{\a}{2} \sin^2\th}=0~,\\
&&\ddot\th+ \frac{m^2}{2}\sin^4\frac{\a}{2} \sin2\th =0~,
\eea
where the first equation can be integrated to obtain the Virasoro constraint
\be\la{xi2a}
q \da^2+\dth^2+ m^2 \prt{\frac{ q}{4}\sin^2 \a+\sin^4\frac{\alpha}{2} \sin^2\th} -4 \k^2 b^2 =0~.
\ee

\textbf{Case I: The $\theta=0$ plane}

We may choose to localise the string solution in the $\th=0$ plane. The motion can be mapped to a solution of the undeformed theory with a different energy. The deformation parameters are absorbed in the energy of the solution by redefining it as
\be
\tilde{\k}=\frac{\k b}{\sqrt{q}}~
\ee
and the motion along the angle $\alpha$ is given by the inverse of the elliptic integral of first kind. Nevertheless, the solution we obtain by departing from $\th=0$ can not be related to the undeformed system. By varying the solution in the invariant plane $\th=p_\th=0$ of the effective Hamiltonian
we get
\be\la{puzzle1}
\eta''(z)\prt{16\k^2+m^2 q( z^2-1)}(z^2-1)+\eta'(z) 2 z\prt{8\k^2+m^2 q(z^2-1)}-m^2 q(1-z)^2 \eta(z)=0~,
\ee
where we have expressed everything in terms of $z=\cos(\a(\t))$ to get a differential equation with rational coefficients.
It turns out that the NVE does not have Liouvillian solutions for any value of $q$, even for $q=1$. We expect that string motion on $\CPthree$ is integrable, and indeed we have performed the analogous analysis for several string solutions on round $\CPthree$ (expressed, for instance, using the metrics (\ref{FSmetricangles}) or (\ref{roundmetric1})). As expected, no signs of non-integrability were found.

To understand this puzzling behaviour, it is useful to re-express the specific string ansatz
above in one of these other coordinate systems, which of course can only be done for $q=1$. In appendix
\ref{FBtoinst} it is shown how to convert to the coordinate
system (\ref{roundmetric1}). Examining the equations of motion for this ansatz in the new coordinate system, we find that there
are three non-trivial equations of motion instead of two in our original coordinate system, for our two functions
$(\theta(\tau),\alpha(\tau))$.\footnote{This is unlike, for instance, the $\CPtwo$ ansatz of the previous section which, when converted to the coordinate
  system (\ref{roundmetric1}) using (\ref{CP2convert}) leads to just two non-trivial equations, whose NVE analysis is consistent with integrability.} It thus appears that, after converting to the round-type
coordinate system (\ref{roundmetric1}), the restriction of string motion to the space (\ref{effectiveansatz1})
has certain peculiarities, at least for $q=1$. It would be very interesting to study this further and understand,
directly in the original coordinate system, whether there is indeed an inconsistency or instability, and whether it
affects only the case $q=1$ (where we are able to map the string to the round $\CPthree$ metric) or the whole solution.

We point out that, as usual, integrability is recovered for point-like particle motion.

\textbf{Case II: The $\theta=\frac{\pi}{2}$ plane}

To elaborate further on the puzzling behavior above we can also localise the string at the equator
of the sphere $\th=\frac{\pi}{2}.$
The equation of motion for $\th$ is satisfied and we remain with the equation of $\a$ \eq{xi1a} or its equivalent \eq{xi2a}. To find the NVE we perturb along the solution $\th=\frac{\pi}{2}+\eta\prt{t}$ to get
\bea
&&\eta''(z)\prt{16\k^2+m^2(z-1)\prt{1+q+(q-1)z}(z^2-1)} \\ \nn
&&+\eta'(z)\prt{16 \k^2 z +m^2(z-1)\prt{1+z+2q z+2(q-1)z^2}}+m^2 q (z-1)^2\eta(z)=0~.
\eea
This equation again does not admit Liouvillian solutions for any $q$ as in \eq{puzzle1} giving the same puzzling behavior for the same reasons discussed there.

\section{Conclusions and Outlook}

In this work we considered two classes of AdS/CFT backgrounds dual to gauge theories with backreacted
flavors. We have shown that the corresponding theories at zero temperature are not integrable, at least in the sector dual to semiclassical strings, and we have studied the existence of chaos for string motion on these backgrounds.

In the case of four-dimensional theories with $\Ncal=2$ supersymmetry, corresponding to the D3/D7 gravity backreacted gravity background, we have found evidence that shows the non-integrability of the theory.  We find analytically closed string solutions that are not integrable, with the integrability restored in the point-like limit for particle motion or when we switch off the backreaction of the D7-flavor branes. This is also the case in the backreacted flavor solution of the D6-branes in the ABJM theory. We conclude that the backreacted flavor deformation breaks the integrability present in the three and four dimensional theories. We have also discussed a puzzling behavior where for a certain string ansatz we did not find Liouvillian solutions of the corresponding NVE even when switching off the backreaction, where the background is just $\CPthree$. As discussed in section \ref{S2localised}, although this puzzling string solution appears valid in our original coordinate system (which is adapted to introducing backreaction), it has some unwelcome features after mapping it to a different metric on $\CPthree$. So we interpret the non-integrable behaviour at $q=1$ as an inconsistency of our string ansatz, and not as a sign of non-integrability of the background. It would be interesting to further study this issue, as it could provide insight into additional (and not previously noticed in the literature) requirements on string solutions, in order for the analytic non-integrability method to be applicable.

In this work we considered two backgrounds with backreacted flavor, one with localised flavor branes and one with the flavor branes smeared along the internal direction. Our belief is that we should expect to see similar behaviours for an even wider class of theories where the virtual quark loops are taken into account and contribute to the gauge propagators. Integrability for these theories should be lost. A strong argument, is that the Hamiltonian of the strings in our methods written in the phase space coordinates capturing the information of non-integrability and chaos, is expected always to be involved enough in the case of backreacting flavor geometries. Given the sensibility of non-integrability and chaos to the Hamiltonian, and in direct analogy to mechanical systems, we can expect that integrability is lost in the presence of unquenched flavors for generic theories.

An important portion of our work consists of analyzing the presence of chaos in the flavor backreacted theories. We have quantified the chaotic motion by computing the leading Lyapunov exponent and observing its dependence on  the $N_c, N_f$ and 't Hooft parameters of the theory. Interestingly we find that the Lyapunov exponent in the  Veneziano regime converges to a certain value which depends on the energy density of the system and the parameters of the theory, but depends weakly on the number of flavors.  We then reduce the number of flavors and, being a bit optimistic
regarding the range of validity of the backreacted solution, we see that the maximum value of the Lyapunov exponent is obtained very quickly as we increase the flavors. Independent phenomena related to properties of the bound states of heavy quarks have been found to be weakly dependent on the numbers of flavors or colors beyond a certain low number,
exactly like the strength of chaos we studied here. It would be interesting if one could establish any connection or common explanation for those phenomena and our observations regarding the chaos. We also note that our analysis is done in zero temperature and the gravity dual theory has no black hole horizon.

We should emphasise that although both of the cases we consider are very good models for backreacted flavor backgrounds and have been used widely in the literature, neither of them is exactly dual to a gauge theory with unquenched flavor: In the D3/D7 case we work in the ``near-core'' region, where the background admits a tractable analytic solution. In the D2/D6 case we work with the background of \cite{Conde:2011sw} which was obtained by considering a smeared distribution of D6 branes instead of a localised one. This leads to a simple squashed $\CPthree$ internal space instead of the much more involved tri-Sasaki space of the localised case, and an associated reduction of supersymmetry from $\Ncal=3$ to $\Ncal=1$. In both cases the full, localised backgrounds are significantly more complicated and it is natural to expect that the loss of integrability seen in our simpler cases
will be a feature of those models as well. In general it is interesting to ask to what extent the analytic non-integrability method, applied to non-exactly known dual backgrounds, can reliably capture the physics of the full gauge theory. Our work can be considered as providing two such examples, which can possibly be validated by further future work going beyond the approximations discussed above.

Our study of the unquenched ABJM theory can act as a starting point for several other studies. For special values of the parameter that controls the squashing of the space and the number of flavors, the background appears in other gauge/gravity dualities.  The backreaction of the $N_f$ flavor D6-branes, smeared over a six-dimensional nearly K\"{a}hler manifold on the $N_c$ color branes, flows to an $\AdS_4$ fixed point dual to Chern-Simons matter theory and which is the special case of our metric \cite{Faedo:2015ula}.  The special case $q=5$ corresponds to an $\cN=1$ supergravity background, being a gravity dual of superconformal Chern-Simons matter theory with $\SO\prt{5}\times \Urm\prt{1}$ global symmetry \cite{Ooguri:2008dk}.

It should also be mentioned that a full study of analytic non-integrability in flavor backgrounds (whether backreacted or not) should also include the open-string sector,  dual to gauge theory operators with fundamental fields at the endpoints instead of traces of adjoint fields. The analytic non-integrability studies performed so far in the literature have been based on closed string solutions, and extending the method to open strings would open up several new avenues for the study of non-integrability. Beyond theories with flavors, this might also include the sector of open strings attached to giant graviton operators, as well as defect CFTs, where the study of integrability was recently initiated in \cite{deLeeuw:2015hxa}. It would be interesting to apply our methods to this sector.

\subsection*{Acknowledgments}

We are thankful to C.S. Chu, K. Hashimoto, C. Kristjansen, K. Sfetsos and K. Yoshida for useful
conversations and correspondence.
This work of DG is supported in part by  the National Center of Theoretical Science
(NCTS) and the grants  101-2112-M-007-021-MY3 and 104-2112-M-007 -001 -MY3 of the
Ministry of Science and Technology of Taiwan. KZ acknowledges support from the National Research Foundation
of South Africa under grant CSUR 93735.
We are also grateful to the South African National Institute for Theoretical Physics (NITheP) for supporting a visit by DG to the University of Pretoria and to NCTS for supporting a visit of KZ to NCTS, Hsinchu. DG would like to thank the Albert Einstein Institute for Fundamental Physics,
Bern University, for hospitality during the final stages of this project. KZ would like to thank the Niels Bohr Institute, Copenhagen, for hospitality and the opportunity to present a preliminary version of this work.

\begin{appendices}

\section{The geometry of $\CPthree$}

The internal part of the metric of \cite{Conde:2011sw}  for generic values of the deformation parameter $q$,  is that
of a squashed $\CPthree$. This is a coset $\Sp(2)/\Urm(2)$. In this appendix we will review some aspects of this
geometry, focusing also on the limit $q=1$ where the metric becomes just $\CPthree$.

\subsection{The squashed $\CPthree$ metric}

Let us start from the metric describing an $\SU(2)$ instanton on $\Srm^4$:
\be \label{S2S4fibremetric}
\diff s^2=q \frac{\diff w_a \diff \wb^a}{(1+w_a\wb^a)^2}+\frac{(\diff u+\Acal)(\diff \bar{u}+\bar{\Acal})}{(1+u\bar{u})^2}~,
\ee
where the $w_a,\wb^a$ are coordinates on the $\Srm^4$ and $u$ is the projective coordinate on $\Srm^2$ (thought of as
a $\CPone$). $\Acal$ is an $\SU(2)$ instanton
given by
\be
\Acal=\frac{(\wb_1+w_2u)(u\diff w_1-\diff \wb_2)+(\wb_2-w_1 u)(u\diff w_2+\diff \wb_1)}{1+w_a\wb^a}\;,
\ee
while $q$ is the squashing parameter which controls the length of the fibre.\footnote{We note that in most of the
literature the squashing parameter is placed
in front of the second term in (\ref{S2S4fibremetric}) and denoted $\lambda^2$. Therefore, effectively our $q$ corresponds
to $1/\lambda^2$.} It is known (e.g. \cite{Aldazabal:2007sn}) that there are two special
values of $q$, which are distinguished by the metric being Einstein: At $q=1$ the metric is equivalent to the
standard Fubini-Study metric on $\CPthree$, which is of course K\"ahler. At $q=2$, the metric is instead
\emph{nearly-K\"ahler}. For all other values of $q$ the metric is not Einstein.

To bring (\ref{S2S4fibremetric}) into the form of the flavor backreacted metric of \cite{Conde:2011sw}, we can start
by choosing coordinates for the $\Srm^4$ part as
\be \label{S4coords}
w_1=\tan\frac{\a}{2} e^{i\frac{\psi_1+\varphi_1}2}\cos\frac{\theta_1}{2}\;,\quad
w_2=-\tan\frac{\a}{2} e^{i\frac{\psi_1-\varphi_1}2}\sin\frac{\theta_1}{2}~.
\ee
Then we find that
\be
\diff_{\Srm^4}^2=\frac{1}{4}\prt{\diff\a^2+\frac{1}{4}\sin^2\a \sum_i\omega_i^2}~,
\ee
where the $\SU(2)$ left-invariant forms we use were given in (\ref{leftinv1}-\ref{leftinv3}).
Turning to the $\Srm^2$ part, we express
\be \label{S2coords}
u=-i\cot\frac{\theta}{2} e^{-i\varphi}\;
\ee
such that
\be
\frac{\diff u\diff \bar{u}}{(1+u\bar{u})^2}=\frac{1}{4}\left(\diff\theta^2+\sin^2\theta\diff\varphi^2\right)~.
\ee
Combining these substitutions we can straightforwardly check that (\ref{S2S4fibremetric}) takes the form
\be
\diff s^2_{\CPthree}=\frac{1}{4}\prt{q\left(\diff \a^2+\frac{1}{4}\sin^2\a \sum_i\omega_i^2\right)+E_1^2+E_2^2}~,
\ee
with the $E_i$ defined in Section \ref{Section:ABJM}. We see that the flavor-backreacted metric of
\cite{Conde:2011sw} is equivalent to the squashed $\CPthree$ metric. It should be noted,
however, that the physics of the backreacted flavor construction restricts $q$ to take the values discussed
below (\ref{qvalues}), which do not include the nearly-K\"ahler case $q=2$.

As mentioned, taking $q=1$ in (\ref{S2S4fibremetric}) one obtains undeformed $\CPthree$, with the instanton-type
metric that appears, for instance, in the twistor construction. We will call this the ``twistor metric'' for $\CPthree$.
Its explicit symmetries are those of the coset $\Sp(2)/\Urm(2)$. It will be useful to compare it with the ``round'' $\SU(4)/\Urm(3)$ metric on
$\CPthree$, which we will do in the following.

\subsection{The Fubini-Study metric on $\CPthree$}

We start by reviewing the Fubini-Study metric on $\CPthree$, which is the one that has been
used in studies of the ABJM geometry in the literature.

Since $\CPthree$ is a K\"ahler manifold, a metric on it can be found through a potential:
\be
\diff s^2 = \del \delb  K = \half g_{i\jbar} \left( \diff z^i\otimes \diff \zb^{\jbar}+\diff\zb^{\jbar}\otimes \diff z^i\right)~,
\ee
with
\be
g_{i\jbar}=\frac{\partial^2}{\partial z^i \partial \zb^{\jbar}} K~,
\ee
where $z, \zb$ are complex coordinates on the manifold.

The K\"ahler potential can be expressed in terms of inhomogeneous coordinates $Z^I$, $I=1\ldots4$, as
\be
K=\ln \bar{Z}^I Z^I~.
\ee
This makes explicit the $\SU(4)$ symmetry of $\CPthree$. However, it is usually convenient to rewrite the metric
in terms of homogeneous coordinates $z^i$, $i=1\ldots 3$, defined as $z^1=Z^2/Z^1,z^2=Z^3/Z^1,z^3=Z^4/Z^1$,
so that $K$ becomes:
\be
K=\ln\left(1+\bar{z}^i z^i\right)~.
\ee
Starting from this potential one obtains the Fubini-Study metric as
\be\label{metricz}
\diff s^2_{\CPthree}=\frac{\diff z^i\diff \zb^i}{1+z^k\zb^k}-\frac{(\zb^i\diff z^i) (z^j\diff \zb^j)}{(1+z^k\zb^k)^2}~,
\ee
while the K\"ahler form is given by
\be \label{Kahlerform}
\omega=\frac{i}{1+\zb^k z^k}\diff z^i\wedge \diff \zb^i-\frac{i}{(1+\zb^k z^k)^2} \zb^i \diff z^i\wedge z^j\diff\zb^j~.
\ee
For applications to string motion, it is convenient to parametrise these coordinates in terms of angles. One commonly
used parametrisation (see e.g. \cite{Pope:1984bd}) is
\be
\begin{split}
z^1&=\tan \frac{\th}{2}\sin\a\sin\frac{\th_1}{2}e^{i \frac{\prt{\psi -\chi}}{2}}e^{i\phi}~,\\
z^2&=\tan \frac{\th}{2}\sin\a\cos\frac{\th_1}{2}e^{i \frac{\prt{\psi +\chi}}{2}}e^{i\phi}~,\\
z^3&=\tan \frac{\th}{2}\cos\a e^{i\phi}~,
\end{split}
\ee
where $0\leq\th,\th_1\leq \pi,~ 0\leq \phi,\chi\leq 2 \pi,~ 0\leq\a\leq\frac{\pi}{2},~ 0\leq\psi\leq 4 \pi$. The metric becomes
\be \label{FSmetricangles}
\diff s^2=\sin^2 \frac{\th}{2}\prt{\diff \a^2+\frac{1}{4}\sin^2\a\prt{\o_1^2+\o_2^2+ \cos^2\a \o_3^2}}+\frac{1}{4}\prt{\diff\th^2 +\sin^2\th\prt{\diff\phi+\frac{1}{2}\sin^2\a \o_3}^2}~,
\ee
where the $\o_i$ are as in (\ref{leftinv1}) with the substitutions $\psi_1\rightarrow\psi$, $\phi_1\rightarrow \chi$.

Another frequently used metric can be obtained from (\ref{metricz}) through the substitution (e.g. \cite{Abbott:2008qd})
\be \label{FStoround1}
z^1=\cot\frac{\theta_2}{2} e^{i\phi_2}\;,
\;\;z^2=\frac{\cot\xi\cos\frac{\theta_1}{2}e^{-\frac{i}{2}(\psi-\phi_1-\phi_2)}}{\sin\frac{\theta_2}{2}}\;,
\;\;z^3=\frac{\cot\xi\sin\frac{\theta_1}{2}e^{-\frac{i}{2}(\psi+\phi_1-\phi_2)}}{\sin\frac{\theta_2}{2}}\;,\;\;
\ee
leading to
\be \label{roundmetric1}
\begin{split}
\diff s^2&=\diff \xi^2+\frac14\cos^2\xi(\diff\theta_1^2+\sin^2\theta_1\diff\phi_1^2)
+\frac14\sin^2\xi(\diff\theta_2^2+\sin^2\theta_2\diff\phi_2^2)\\
&+\frac14\sin^2\xi\cos^2\xi(\diff\psi-\cos\theta_1\diff\phi_1+\cos\theta_1\diff\phi_2)^2~.
\end{split}
\ee
This parametrisation is used e.g. in \cite{Cvetic:2000yp}. In the next section we show how some of our anz\"atze for string motion, written in the twistor-like coordinate
system which is appropriate for squashing, can be expressed in this metric, and we will also make use of it in Appendix \ref{T11restriction}, where we study
integrability on a $T^{11}$ subspace of $\CPthree$.

\subsection{From the Fubini-Study metric to the twistor metric} \label{FBtoinst}

The relation between the Fubini-Study metric (\ref{metricz}) and the twistor metric ((\ref{S2S4fibremetric}) with $q=1$)
has been discussed in \cite{Bellucci:2003yx} in the context of relating the Hall mechanics of a particle on $\Srm^4$ and $\CPthree$ (see also \cite{Krivonostalk} for more details and generalisations). The relation is simply the following:
\be \label{redef}
z^1=w_1 u -\wb_2\;,\quad z^2=w_2 u+\wb_1\;,\quad z^3=u~.
\ee
Performing this coordinate change and substituting (\ref{S4coords}) and (\ref{S2coords}), it can be checked that
one obtains the twistor metric, while, as also mentioned in \cite{Conde:2011sw}, the K\"ahler form
(\ref{Kahlerform}) can be expressed in terms of the $E^{1,2}, \Scal^\alpha$ and $\Scal^i$ forms defined in Section \ref{Section:ABJM} as
\be
\omega=\half\left(E^1\wedge E^2-\Scal^\alpha\wedge \Scal^3-\Scal^1\wedge \Scal^2\right)\;.
\ee
Of course, in order to compare explicit string solutions we would like to go beyond the identification (\ref{redef})
and relate angle coordinates for the Fubini-Study metric to those for the twistor metric. Although this is not
straightforward to do in full generality, it can be done if we restrict to specific  subspaces of
$\CPthree$.

One such case concerns the ansatz discussed in section \ref{S2localised}, which we repeat
here for convenience:
\be
\alpha=\alpha(\tau)\;,\quad\theta=0\;,\quad\phi_1=m\sigma\;,\quad\phi=\frac{\pi}2\;,\quad\theta_1=\theta_1(\tau)\;,\quad\psi_1=\frac{\pi}2 \;.
\ee
Combining the expressions  (\ref{S4coords}), (\ref{S2coords}) and (\ref{redef}) we find
that this corresponds to the following choice for the $z^i$ coordinates:
\be \label{ziansatz3}
z^1=-i\,e^{{{im\sigma}\over{2}}}\,\tan\frac{\alpha(\tau)}{2}\cot\frac{\theta(\tau)}2\;,\quad
z^2=e^{-\frac{i m \sigma}{2}}\tan\frac{\alpha(\tau)}{2}\;,\quad z^3=-i \cos\frac{\theta(\tau)}{2}~.
\ee
Let us now convert the above parametrisation to the coordinate system (\ref{roundmetric1}), which is commonly used in
studies of $\CPthree$.  To do so, we make the following identification of coordinates:
\be
\xi(\t)=\frac{\alpha(\tau)-\pi}{2},\quad\theta_1(\tau)=\theta(\tau),\quad \theta_2(\tau)=\theta(\tau),\quad \phi_1=m\sigma+\frac{3\pi}2,\quad
\phi_2=-\frac{\pi}{2},\quad\psi=0\;.
\ee
With this choice, and also redefining (\ref{FStoround1}) as $z^1\rightarrow z^3, z^2\rightarrow z^2, z^3\rightarrow z^2$,
the inhomogeneous coordinates (\ref{FStoround1}) become equal to (\ref{ziansatz3}).

For completeness, let us also mention how the $\CPtwo$ ansatz (\ref{CP2ansatz}) can be written in the coordinates of (\ref{roundmetric1}). One redefines
the $z^i$ as above, this time with the following identifications:
\be \label{CP2convert}
\xi(\t)=\frac{\alpha(\tau)-\pi}{2}\;,\quad \theta_1=\theta_2=\frac{\pi}{2}\;,\quad \phi_1=m\sigma-\frac{\pi}2\;,\quad\phi_2=\pi,\quad \psi(\t)=-\theta_1(\tau)\;.
\ee

\subsection{Restriction to $\CPtwo$} \label{CP2restriction}

As noted in \cite{Conde:2011sw}, an interesting subspace of the $\Srm^2$-bundle
over $\Srm^4$ that we have been discussing arises when the two-sphere angles take the values
\be
\th=\frac{\pi}{2}~,\quad \varphi=\frac{\pi}{2}~,
\ee
giving the four-dimensional metric
\be \label{CP2qnot1}
ds^2= \frac{1}{4}\prt{q d\a^2+ \frac{1}{2}\prt{1+q+(1-q)\cos\a}\sin^2 \frac{\a}{2}\prt{\o_1^2+\o_3^2}+\frac{q \sin^2\a}{4} \o_2^2}\;.
\ee
If we turn off the backreaction by setting $q=1$, this metric is
\be
ds^2_{CP^2}= \frac{1}{4}\prt{d\a^2+ \sin^2 \frac{\a}{2}\prt{\o_1^2+\o_3^2}+\frac{\sin^2\a}{4} \o_2^2}~.
\ee
This is the Fubini-Study metric on $\CPtwo$. As string motion on $\CPtwo$ is expected to be integrable, it is
worthwhile to perform an analysis of integrability on this subspace, for $q\neq 1$ as well.
This is addressed in section \ref{CP2section}.

\subsection{Seven-sphere lift}

The family of squashed $\CPthree$ metrics is also relevant for the construction of squashed seven-sphere
metrics, which are $\Srm^1$ bundles over a squashed $\CPthree$ base. The round metric on $\Srm^7$ is equivalent to
the metric obtained via an $\Srm^1$
bundle over the $q=1$ $\CPthree$ metric. There is one other value of $q$ for which the squashed $\Srm^7$
space is Einstein, which is $q=5$: Even though the squashed $\CPthree$ metric is not Einstein for this
value, the corresponding squashed seven-sphere is. The AdS/CFT dual theory to this squashed sphere
has been considered in \cite{Ooguri:2008dk}.

 Let us briefly discuss the uplift of $\CPthree$ (for $q=1$) to $\Srm^7$, following the discussion in
\cite{Conde:2011sw}. Calling the additional direction $\psi$, and
recalling that $\theta$ and $\varphi$ are the $\Srm^2$ angles, we introduce a new set of left-invariant forms,
\be
\begin{split}
\tilde{\omega}^1&=-\sin\varphi\diff\theta+\cos\varphi\sin\theta\diff\psi~,\\
\tilde{\omega}^2&=\cos\varphi\diff \theta+\sin\varphi\sin\theta\diff \psi~,\\
\tilde{\omega}^3&=\diff \varphi+\cos\theta\diff\psi~,
\end{split}
\ee
in terms of which the $\Srm^7$ metric is\footnote{For the actual uplift of ABJM to M-theory one also needs to perform a $\mathbb{Z}_k$
orbifold of the $\psi$ angle.}
\be \label{Hopfmetric}
\diff s^2 =\frac{1}{4}\left(\diff s_{\Srm^4}^2+\sum_{i=1}^3 \left(\tilde{\omega}^i+A^i\right)^2\right)~,
\ee
where we recall that $A^i=-\sin^2\frac{\a}{2}\omega^i=-\frac{1-\cos\alpha}{2}\omega^i$. We see that the uplift of the twistor metric of $\CPthree$
naturally exhibits $\Srm^7$ as an $\Srm^3$ (Hopf) fibration over $\Srm^4$.

Similarly to the discussion above for $\CPthree$, there is a question of how to show the equivalence of the above
$\Srm^7$ metric to the more usual round  metric on $\Srm^7$. This is addressed e.g. in \cite{Hatsuda:2009vj}. For
completeness we sketch their construction below, while referring to their work for additional details. Let us start by
writing (\ref{Hopfmetric}) more explicitly (also suppressing the summation symbol) as
\be
\diff s^2=\frac{1}{4}\diff\alpha^2+\frac{1}{16}\sin^2\alpha \omega_i^2+\frac{1}{4}(\tilde{\omega}_i+A_i)^2~,
\ee
in order to better exhibit the roles of the two sets of left-invariant forms.
Let us now write
\be \label{omegashift}
\omega_i=\omega_i-\tilde{\omega}_i+\tilde{\omega}_i=\nu_i+\tilde{\omega}_i~,
\ee
i.e. we have defined a new one-form $\nu_i=\omega_i-\tilde{\omega}_i$. Then
\be
\frac{\sin^2\alpha}{16}\omega_i^2=\frac{\sin^2\alpha}{16}(\nu_i^2+2 \nu_i \tilde{\omega}_i +\tilde{\omega}_i^2)~,
\ee
while
\be
(\tilde{\omega}_i-\frac{1-\cos\alpha}{2}\omega_i)^2=\cos\alpha \tilde{\omega}_i^2-(1-\cos\alpha)\tilde{\omega}_i\nu_i
+\frac{(1-\cos\alpha)^2}{4}(\nu_i^2+2\nu_i \tilde{\omega}_i +\tilde{\omega}_i^2)~.
\ee
Adding these terms, and also substituting $\alpha=2\hat{\alpha}$, a short calculation leads to
\be
\frac{1}{4}\diff\alpha^2+\frac{1}{16}\sin^2\alpha \omega_i^2+\frac{1}{4}(\tilde{\omega}_i+A_i)^2
=\diff\hat{\alpha}^2+\frac14\sin^2\hat{\alpha}~ \nu_i^2+\frac14\cos^2\hat{\alpha} ~\tilde{\omega}_i^2~.
\ee
The right-hand side is what is commonly considered as the round metric on $\Srm^7$. For more details on how
this relation between the metrics is derived, we refer to \cite{Hatsuda:2009vj}. Of course, although (\ref{omegashift})
is suggestive, the actual identification of angles that implements this transformation is not straightforward.

\section{Strings on the $T^{11}$ restriction of $\CPthree$} \label{T11restriction}

 An interesting feature of the round $\CPthree$ metric is that it admits a restriction to a 5-dimensional $T^{11}$
 space. This can be most clearly seen starting from the metric (\ref{roundmetric1}) \cite{Cvetic:2000yp}. The constant $\xi$
 surfaces are $\Urm(1)$ bundles on $\Srm^2\times \Srm^2$, and as such the restriction of (\ref{roundmetric1}) to constant $\xi$
 has the geometric features of $T^{11}$. However, the metric one obtains by taking $\xi$ constant is not the Einstein metric on $T^{11}$,
 which requires relations between the coefficients of the three factors that cannot be satisfied for any choice of $\xi$.

 Integrability of string motion on $T^{11}$ (and its generalisations with arbitrary coefficients for
the three factors) was considered in \cite{Basu:2011fw}, where it was established that these backgrounds
are not integrable. So one can wonder how they can arise as foliations of the integrable $\CPthree$ geometry.
We can examine this question by considering string motion on  $\CPthree$, parametrised as in (\ref{roundmetric1}), and
choosing an ansatz for a string that lives at constant $\xi$:
\be
\begin{split}
t(\tau,\sigma)&=k\tau\;,\quad\xi(\tau,\sigma)=\xi\;,\quad\theta_1(\tau,\sigma)=\theta_1(\tau)\;,\quad
\phi_1(\tau,\sigma)=m\sigma\;,\;\\
&\theta_2(\tau,\sigma)=\theta_2(\tau)\;,\quad \phi_2(\tau,\sigma)=n\sigma\;,\quad
\psi(\tau,\sigma)=0~.
\end{split}
\ee
While most equations of motion following from this ansatz are equivalent to those of \cite{Basu:2011fw},
there is an additional equation coming from the $\xi$ equation of motion:
\be
\dot{\theta}_1^2-\dot{\theta}_2^2+(\cos^2\xi-\sin^2\xi)(m\cos\theta_1-n\cos\theta_2)^2 -m^2\sin^2\theta_1+n^2\sin\theta_2~.
\ee
Following \cite{Basu:2011fw}, we take our straight line solution to be $\theta_2=0$. It turns out
that the additional equation of motion then restricts $\theta_1$ to take a constant value:
\be
\bar{\theta}_1=\arccos\left(\frac23\frac{(2\cos^2\xi-1)n}{m\cos^2\xi}\right)~.
\ee
Perturbing $\theta_2(\tau)=0+\eta(\tau)$, linearising and substituting the solution for $\theta_1$,
we find the NVE
\be
\ddot{\eta}+\frac{n^2}{3}(1+\cos^2\xi)=0~.
\ee
This is integrable, unlike the NVE found for $T^{11}$ in \cite{Basu:2011fw} (where $\theta_1$ was
a function of $\tau$). So it appears that the
additional EOM has led us to a simpler NVE which does not exhibit non-integrable behaviour. This is
of course consistent with our expectations from integrability of the $\CPthree$ sigma model.

\end{appendices}



\begin{thebibliography}{10}

\bibitem{Beisert:2010jr}
N.~Beisert {\em et~al.}, ``{Review of AdS/CFT Integrability: An Overview},''
  \href{http://dx.doi.org/10.1007/s11005-011-0529-2}{{\em Lett. Math. Phys.}
  {\bfseries 99} (2012) 3--32},
\href{http://arxiv.org/abs/1012.3982}{{\ttfamily arXiv:1012.3982 [hep-th]}}.

\bibitem{Bombardelli:2016rwb}
D.~Bombardelli, A.~Cagnazzo, R.~Frassek, F.~Levkovich-Maslyuk, F.~Loebbert,
  S.~Negro, I.~M. Szecsenyi, A.~Sfondrini, S.~J. van Tongeren, and
  A.~Torrielli, ``{An integrability primer for the gauge-gravity
  correspondence: An introduction},''
  \href{http://dx.doi.org/10.1088/1751-8113/49/32/320301}{{\em J. Phys.}
  {\bfseries A49} no.~32, (2016) 320301},
\href{http://arxiv.org/abs/1606.02945}{{\ttfamily arXiv:1606.02945 [hep-th]}}.

\bibitem{Basu:2011fw}
P.~Basu and L.~A. Pando~Zayas, ``{Analytic Non-integrability in String
  Theory},'' \href{http://dx.doi.org/10.1103/PhysRevD.84.046006}{{\em Phys.
  Rev.} {\bfseries D84} (2011) 046006},
\href{http://arxiv.org/abs/1105.2540}{{\ttfamily arXiv:1105.2540 [hep-th]}}.

\bibitem{MoralesRuiz99}
J.~J. Morales-Ruiz, {\em Differential Galois theory and non-integrability of
  Hamiltonian Systems}.
\newblock Birkhauser, Basel, 1999.

\bibitem{Maldacena:2015waa}
J.~Maldacena, S.~H. Shenker, and D.~Stanford, ``{A bound on chaos},''
  \href{http://dx.doi.org/10.1007/JHEP08(2016)106}{{\em JHEP} {\bfseries 08}
  (2016) 106},
\href{http://arxiv.org/abs/1503.01409}{{\ttfamily arXiv:1503.01409 [hep-th]}}.

\bibitem{Shenker:2013pqa}
S.~H. Shenker and D.~Stanford, ``{Black holes and the butterfly effect},''
  \href{http://dx.doi.org/10.1007/JHEP03(2014)067}{{\em JHEP} {\bfseries 03}
  (2014) 067},
\href{http://arxiv.org/abs/1306.0622}{{\ttfamily arXiv:1306.0622 [hep-th]}}.

\bibitem{Shenker:2013yza}
S.~H. Shenker and D.~Stanford, ``{Multiple Shocks},''
  \href{http://dx.doi.org/10.1007/JHEP12(2014)046}{{\em JHEP} {\bfseries 12}
  (2014) 046},
\href{http://arxiv.org/abs/1312.3296}{{\ttfamily arXiv:1312.3296 [hep-th]}}.

\bibitem{Hashimoto:2016dfz}
K.~Hashimoto and N.~Tanahashi, ``{Universality in Chaos of Particle Motion near
  Black Hole Horizon},''
  \href{http://dx.doi.org/10.1103/PhysRevD.95.024007}{{\em Phys. Rev.}
  {\bfseries D95} no.~2, (2017) 024007},
\href{http://arxiv.org/abs/1610.06070}{{\ttfamily arXiv:1610.06070 [hep-th]}}.

\bibitem{Erdmenger:2007cm}
J.~Erdmenger, N.~Evans, I.~Kirsch, and E.~Threlfall, ``{Mesons in Gauge/Gravity
  Duals - A Review},'' \href{http://dx.doi.org/10.1140/epja/i2007-10540-1}{{\em
  Eur. Phys. J.} {\bfseries A35} (2008) 81--133},
\href{http://arxiv.org/abs/0711.4467}{{\ttfamily arXiv:0711.4467 [hep-th]}}.

\bibitem{Nunez:2010sf}
C.~Nunez, A.~Paredes, and A.~V. Ramallo, ``{Unquenched Flavor in the
  Gauge/Gravity Correspondence},''
  \href{http://dx.doi.org/10.1155/2010/196714}{{\em Adv. High Energy Phys.}
  {\bfseries 2010} (2010) 196714},
\href{http://arxiv.org/abs/1002.1088}{{\ttfamily arXiv:1002.1088 [hep-th]}}.


\bibitem{Karch:2002sh}
A.~Karch and E.~Katz, ``{Adding flavor to AdS / CFT},''
  \href{http://dx.doi.org/10.1088/1126-6708/2002/06/043}{{\em JHEP} {\bfseries
  06} (2002) 043},
\href{http://arxiv.org/abs/hep-th/0205236}{{\ttfamily arXiv:hep-th/0205236}}.

\bibitem{Burrington:2004id}
B.~A. Burrington, J.~T. Liu, L.~A. Pando~Zayas, and D.~Vaman, ``{Holographic
  duals of flavored N=1 super Yang-mills: Beyond the probe approximation},''
  \href{http://dx.doi.org/10.1088/1126-6708/2005/02/022}{{\em JHEP} {\bfseries
  02} (2005) 022},
\href{http://arxiv.org/abs/hep-th/0406207}{{\ttfamily arXiv:hep-th/0406207}}.


\bibitem{Kirsch:2005uy}
I.~Kirsch and D.~Vaman, ``{The D3 / D7 background and flavor dependence of
  Regge trajectories},''
  \href{http://dx.doi.org/10.1103/PhysRevD.72.026007}{{\em Phys. Rev.}
  {\bfseries D72} (2005) 026007},
\href{http://arxiv.org/abs/hep-th/0505164}{{\ttfamily arXiv:hep-th/0505164}}.


\bibitem{Aharony:2008ug}
O.~Aharony, O.~Bergman, D.~L. Jafferis, and J.~Maldacena, ``{N=6 superconformal
  Chern-Simons-matter theories, M2-branes and their gravity duals},''
  \href{http://dx.doi.org/10.1088/1126-6708/2008/10/091}{{\em JHEP} {\bfseries
  10} (2008) 091},
\href{http://arxiv.org/abs/0806.1218}{{\ttfamily arXiv:0806.1218 [hep-th]}}.


\bibitem{Conde:2011sw}
E.~Conde and A.~V. Ramallo, ``{On the gravity dual of Chern-Simons-matter
  theories with unquenched flavor},''
  \href{http://dx.doi.org/10.1007/JHEP07(2011)099}{{\em JHEP} {\bfseries 1107}
  (2011) 099},
\href{http://arxiv.org/abs/1105.6045}{{\ttfamily arXiv:1105.6045 [hep-th]}}.

\bibitem{Bea:2013jxa}
Y.~Bea, E.~Conde, N.~Jokela, and A.~V. Ramallo, ``{Unquenched massive flavors
  and flows in Chern-Simons matter theories},''
  \href{http://dx.doi.org/10.1007/JHEP12(2013)033}{{\em JHEP} {\bfseries 12}
  (2013) 033},
\href{http://arxiv.org/abs/1309.4453}{{\ttfamily arXiv:1309.4453 [hep-th]}}.

\bibitem{Faedo:2015ula}
A.~F. Faedo, D.~Mateos, and J.~Tarrio, ``{Three-dimensional super Yang-Mills
  with unquenched flavor},''
  \href{http://dx.doi.org/10.1007/JHEP07(2015)056}{{\em JHEP} {\bfseries 07}
  (2015) 056},
\href{http://arxiv.org/abs/1505.00210}{{\ttfamily arXiv:1505.00210 [hep-th]}}.

\bibitem{Ooguri:2008dk}
H.~Ooguri and C.-S. Park, ``{Superconformal Chern-Simons Theories and the
  Squashed Seven Sphere},''
  \href{http://dx.doi.org/10.1088/1126-6708/2008/11/082}{{\em JHEP} {\bfseries
  11} (2008) 082},
\href{http://arxiv.org/abs/0808.0500}{{\ttfamily arXiv:0808.0500 [hep-th]}}.

\bibitem{Hashimoto:2016wme}
K.~Hashimoto, K.~Murata, and K.~Yoshida, ``{Chaos of chiral condensate},''
 Phys.\ Rev.\ Lett.\  {\bf 117} (2016) no.23,  231602,
\href{http://arxiv.org/abs/1605.08124}{{\ttfamily arXiv:1605.08124 [hep-th]}}.

\bibitem{Stepanchuk:2012xi}
A.~Stepanchuk and A.~A. Tseytlin, ``{On (non)integrability of classical strings
  in p-brane backgrounds},''
  \href{http://dx.doi.org/10.1088/1751-8113/46/12/125401}{{\em J. Phys.}
  {\bfseries A46} (2013) 125401},
\href{http://arxiv.org/abs/1211.3727}{{\ttfamily arXiv:1211.3727 [hep-th]}}.

\bibitem{Giataganas:2013dha}
D.~Giataganas, L.~A. Pando~Zayas, and K.~Zoubos, ``{On Marginal Deformations
  and Non-Integrability},''
  \href{http://dx.doi.org/10.1007/JHEP01(2014)129}{{\em JHEP} {\bfseries 1401}
  (2014) 129},
\href{http://arxiv.org/abs/1311.3241}{{\ttfamily arXiv:1311.3241 [hep-th]}}.

\bibitem{Basu:2012ae}
P.~Basu, D.~Das, A.~Ghosh, and L.~A. Pando~Zayas, ``{Chaos around Holographic
  Regge Trajectories},'' \href{http://dx.doi.org/10.1007/JHEP05(2012)077}{{\em
  JHEP} {\bfseries 05} (2012) 077},
\href{http://arxiv.org/abs/1201.5634}{{\ttfamily arXiv:1201.5634 [hep-th]}}.

\bibitem{Ishii:2016rlk}
T.~Ishii, K.~Murata, and K.~Yoshida, ``{Fate of chaotic strings in a confining
  geometry},'' \href{http://dx.doi.org/10.1103/PhysRevD.95.066019}{{\em Phys.
  Rev.} {\bfseries D95} no.~6, (2017) 066019},
\href{http://arxiv.org/abs/1610.05833}{{\ttfamily arXiv:1610.05833 [hep-th]}}.

\bibitem{Chervonyi:2013eja}
Y.~Chervonyi and O.~Lunin, ``{(Non)-Integrability of Geodesics in D-brane
  Backgrounds},'' \href{http://dx.doi.org/10.1007/JHEP02(2014)061}{{\em JHEP}
  {\bfseries 02} (2014) 061},
\href{http://arxiv.org/abs/1311.1521}{{\ttfamily arXiv:1311.1521 [hep-th]}}.

\bibitem{Koch:2011hb}
R.~de~Mello~Koch, M.~Dessein, D.~Giataganas, and C.~Mathwin, ``{Giant Graviton
  Oscillators},'' \href{http://dx.doi.org/10.1007/JHEP10(2011)009}{{\em JHEP}
  {\bfseries 10} (2011) 009},
\href{http://arxiv.org/abs/1108.2761}{{\ttfamily arXiv:1108.2761 [hep-th]}}.

\bibitem{Giataganas:2014hma}
D.~Giataganas and K.~Sfetsos, ``{Non-integrability in non-relativistic
  theories},'' \href{http://dx.doi.org/10.1007/JHEP06(2014)018}{{\em JHEP}
  {\bfseries 06} (2014) 018},
\href{http://arxiv.org/abs/1403.2703}{{\ttfamily arXiv:1403.2703 [hep-th]}}.

\bibitem{Farahi:2014lta}
A.~Farahi and L.~A. Pando~Zayas, ``{Gravitational Collapse, Chaos in CFT
  Correlators and the Information Paradox},''
  \href{http://dx.doi.org/10.1016/j.physletb.2014.05.017}{{\em Phys. Lett.}
  {\bfseries B734} (2014) 31--35},
\href{http://arxiv.org/abs/1402.3592}{{\ttfamily arXiv:1402.3592 [hep-th]}}.

\bibitem{Ma:2014aha}
D.-Z. Ma, J.-P. Wu, and J.~Zhang, ``{Chaos from the ring string in a
  Gauss-Bonnet black hole in AdS5 space},''
  \href{http://dx.doi.org/10.1103/PhysRevD.89.086011}{{\em Phys. Rev.}
  {\bfseries D89} no.~8, (2014) 086011},
\href{http://arxiv.org/abs/1405.3563}{{\ttfamily arXiv:1405.3563 [hep-th]}}.

\bibitem{Asano:2015eha}
Y.~Asano, D.~Kawai, and K.~Yoshida, ``{Chaos in the BMN matrix model},''
  \href{http://dx.doi.org/10.1007/JHEP06(2015)191}{{\em JHEP} {\bfseries 06}
  (2015) 191},
\href{http://arxiv.org/abs/1503.04594}{{\ttfamily arXiv:1503.04594 [hep-th]}}.

\bibitem{Asano:2015qwa}
Y.~Asano, D.~Kawai, H.~Kyono, and K.~Yoshida, ``{Chaotic strings in a near
  Penrose limit of AdS$_{5} \times$ T$^{1,1}$},''
  \href{http://dx.doi.org/10.1007/JHEP08(2015)060}{{\em JHEP} {\bfseries 08}
  (2015) 060},
\href{http://arxiv.org/abs/1505.07583}{{\ttfamily arXiv:1505.07583 [hep-th]}}.

\bibitem{Panigrahi:2016zny}
K.~L. Panigrahi and M.~Samal, ``{Chaos in classical string dynamics in
  $\hat{\gamma}$ deformed $AdS_5 \times T^{1,1}$},''
  \href{http://dx.doi.org/10.1016/j.physletb.2016.08.021}{{\em Phys. Lett.}
  {\bfseries B761} (2016) 475--481},
\href{http://arxiv.org/abs/1605.05638}{{\ttfamily arXiv:1605.05638 [hep-th]}}.

\bibitem{Asano:2016qsv}
Y.~Asano, H.~Kyono, and K.~Yoshida, ``{Melnikov's method in String Theory},''
  \href{http://dx.doi.org/10.1007/JHEP09(2016)103}{{\em JHEP} {\bfseries 09}
  (2016) 103},
\href{http://arxiv.org/abs/1607.07302}{{\ttfamily arXiv:1607.07302 [hep-th]}}.

\bibitem{Basu:2016zkr}
P.~Basu, P.~Chaturvedi, and P.~Samantray, ``{Chaotic dynamics of strings in
  charged black hole backgrounds},''
\href{http://arxiv.org/abs/1607.04466}{{\ttfamily arXiv:1607.04466 [hep-th]}}.

\bibitem{Mann:2006rh}
N.~Mann and S.~E. Vazquez, ``{Classical Open String Integrability},''
  \href{http://dx.doi.org/10.1088/1126-6708/2007/04/065}{{\em JHEP} {\bfseries
  04} (2007) 065},
\href{http://arxiv.org/abs/hep-th/0612038}{{\ttfamily arXiv:hep-th/0612038}}.

\bibitem{Stefanski:2003qr}
B.~Stefanski, Jr., ``{Open spinning strings},''
  \href{http://dx.doi.org/10.1088/1126-6708/2004/03/057}{{\em JHEP} {\bfseries
  03} (2004) 057},
\href{http://arxiv.org/abs/hep-th/0312091}{{\ttfamily arXiv:hep-th/0312091}}.


\bibitem{Chen:2004yf}
B.~Chen, X.-J. Wang, and Y.-S. Wu, ``{Open spin chain and open spinning
  string},'' \href{http://dx.doi.org/10.1016/j.physletb.2004.04.013}{{\em Phys.
  Lett.} {\bfseries B591} (2004) 170--180},
\href{http://arxiv.org/abs/hep-th/0403004}{{\ttfamily arXiv:hep-th/0403004}}.


\bibitem{Aharony:1998xz}
O.~Aharony, A.~Fayyazuddin, and J.~M. Maldacena, ``{The Large N limit of N=2,
  N=1 field theories from three-branes in F theory},''
  \href{http://dx.doi.org/10.1088/1126-6708/1998/07/013}{{\em JHEP} {\bfseries
  07} (1998) 013},
\href{http://arxiv.org/abs/hep-th/9806159}{{\ttfamily arXiv:hep-th/9806159}}.

\bibitem{Grana:2001xn}
M.~Grana and J.~Polchinski, ``{Gauge / gravity duals with holomorphic
  dilaton},'' \href{http://dx.doi.org/10.1103/PhysRevD.65.126005}{{\em Phys.
  Rev.} {\bfseries D65} (2002) 126005},
\href{http://arxiv.org/abs/hep-th/0106014}{{\ttfamily arXiv:hep-th/0106014
  [hep-th]}}.

\bibitem{Gesztesy:1977vd}
F.~Gesztesy and L.~Pittner, ``{Electrons in Logarithmic Potentials. 1. Solution
  of the Schrodinger Equation},''
\href{http://dx.doi.org/10.1088/0305-4470/11/4/008}{{\em J. Phys.} {\bfseries
  A11} (1978) 679}.

\bibitem{Arutyunov:2008if}
G.~Arutyunov and S.~Frolov, ``{Superstrings on AdS(4) x CP**3 as a Coset
  Sigma-model},'' \href{http://dx.doi.org/10.1088/1126-6708/2008/09/129}{{\em
  JHEP} {\bfseries 0809} (2008) 129},
\href{http://arxiv.org/abs/0806.4940}{{\ttfamily arXiv:0806.4940 [hep-th]}}.

\bibitem{Stefanski:2008ik}
B.~Stefanski, jr, ``{Green-Schwarz action for Type IIA strings on AdS(4) x
  CP**3},'' \href{http://dx.doi.org/10.1016/j.nuclphysb.2008.09.015}{{\em Nucl.
  Phys.} {\bfseries B808} (2009) 80--87},
\href{http://arxiv.org/abs/0806.4948}{{\ttfamily arXiv:0806.4948 [hep-th]}}.

\bibitem{Hohenegger:2009as}
S.~Hohenegger and I.~Kirsch, ``{A Note on the holography of Chern-Simons matter
  theories with flavour},''
  \href{http://dx.doi.org/10.1088/1126-6708/2009/04/129}{{\em JHEP} {\bfseries
  04} (2009) 129},
\href{http://arxiv.org/abs/0903.1730}{{\ttfamily arXiv:0903.1730 [hep-th]}}.

\bibitem{Gaiotto:2009tk}
D.~Gaiotto and D.~L. Jafferis, ``{Notes on adding D6 branes wrapping RP**3 in
  AdS(4) x CP**3},'' \href{http://dx.doi.org/10.1007/JHEP11(2012)015}{{\em
  JHEP} {\bfseries 11} (2012) 015},
\href{http://arxiv.org/abs/0903.2175}{{\ttfamily arXiv:0903.2175 [hep-th]}}.

\bibitem{Hikida:2009tp}
Y.~Hikida, W.~Li, and T.~Takayanagi, ``{ABJM with Flavors and FQHE},''
  \href{http://dx.doi.org/10.1088/1126-6708/2009/07/065}{{\em JHEP} {\bfseries
  07} (2009) 065},
\href{http://arxiv.org/abs/0903.2194}{{\ttfamily arXiv:0903.2194 [hep-th]}}.

\bibitem{Jensen:2010vx}
K.~Jensen, ``{More Holographic Berezinskii-Kosterlitz-Thouless Transitions},''
  \href{http://dx.doi.org/10.1103/PhysRevD.82.046005}{{\em Phys. Rev.}
  {\bfseries D82} (2010) 046005},
\href{http://arxiv.org/abs/1006.3066}{{\ttfamily arXiv:1006.3066 [hep-th]}}.

\bibitem{Ammon:2009wc}
M.~Ammon, J.~Erdmenger, R.~Meyer, A.~O'Bannon, and T.~Wrase, ``{Adding Flavor
  to AdS(4)/CFT(3)},''
  \href{http://dx.doi.org/10.1088/1126-6708/2009/11/125}{{\em JHEP} {\bfseries
  11} (2009) 125},
\href{http://arxiv.org/abs/0909.3845}{{\ttfamily arXiv:0909.3845 [hep-th]}}.

\bibitem{Zafrir:2012yg}
G.~Zafrir, ``{Embedding massive flavor in ABJM},''
  \href{http://dx.doi.org/10.1007/JHEP10(2012)056}{{\em JHEP} {\bfseries 10}
  (2012) 056},
\href{http://arxiv.org/abs/1202.4295}{{\ttfamily arXiv:1202.4295 [hep-th]}}.

\bibitem{Bai:2017jpe}
N.~Bai, H.-H. Chen, S.~He, J.-B. Wu, W.-L. Yang, and M.-Q. Zhu, ``{Integrable
  Open Spin Chains from Flavored ABJM Theory},''
\href{http://arxiv.org/abs/1704.05807}{{\ttfamily arXiv:1704.05807 [hep-th]}}.

\bibitem{Bianchi:2009ja}
M.~S. Bianchi, S.~Penati, and M.~Siani, ``{Infrared stability of ABJ-like
  theories},'' \href{http://dx.doi.org/10.1007/JHEP01(2010)080}{{\em JHEP}
  {\bfseries 01} (2010) 080},
\href{http://arxiv.org/abs/0910.5200}{{\ttfamily arXiv:0910.5200 [hep-th]}}.

\bibitem{Bianchi:2009rf}
M.~S. Bianchi, S.~Penati, and M.~Siani, ``{Infrared Stability of N = 2
  Chern-Simons Matter Theories},''
  \href{http://dx.doi.org/10.1007/JHEP05(2010)106}{{\em JHEP} {\bfseries 05}
  (2010) 106},
\href{http://arxiv.org/abs/0912.4282}{{\ttfamily arXiv:0912.4282 [hep-th]}}.

\bibitem{deLeeuw:2015hxa}
M.~de~Leeuw, C.~Kristjansen, and K.~Zarembo, ``{One-point Functions in Defect
  CFT and Integrability},''
  \href{http://dx.doi.org/10.1007/JHEP08(2015)098}{{\em JHEP} {\bfseries 08}
  (2015) 098},
\href{http://arxiv.org/abs/1506.06958}{{\ttfamily arXiv:1506.06958 [hep-th]}}.

\bibitem{Aldazabal:2007sn}
G.~Aldazabal and A.~Font, ``{A Second look at N=1 supersymmetric AdS(4) vacua
  of type IIA supergravity},''
  \href{http://dx.doi.org/10.1088/1126-6708/2008/02/086}{{\em JHEP} {\bfseries
  02} (2008) 086},
\href{http://arxiv.org/abs/0712.1021}{{\ttfamily arXiv:0712.1021 [hep-th]}}.

\bibitem{Pope:1984bd}
C.~N. Pope and N.~P. Warner, ``{An SU(4) Invariant Compactification of $d=11$
  Supergravity on a Stretched Seven Sphere},''
\href{http://dx.doi.org/10.1016/0370-2693(85)90992-X}{{\em Phys. Lett.}
  {\bfseries B150} (1985) 352--356}.

\bibitem{Abbott:2008qd}
M.~C. Abbott and I.~Aniceto, ``{Giant Magnons in AdS(4) x CP**3: Embeddings,
  Charges and a Hamiltonian},''
  \href{http://dx.doi.org/10.1088/1126-6708/2009/04/136}{{\em JHEP} {\bfseries
  04} (2009) 136},
\href{http://arxiv.org/abs/0811.2423}{{\ttfamily arXiv:0811.2423 [hep-th]}}.

\bibitem{Cvetic:2000yp}
M.~Cvetic, H.~Lu, and C.~N. Pope, ``{Consistent warped space Kaluza-Klein
  reductions, half maximal gauged supergravities and CP**n constructions},''
  \href{http://dx.doi.org/10.1016/S0550-3213(00)00708-2}{{\em Nucl. Phys.}
  {\bfseries B597} (2001) 172--196},
\href{http://arxiv.org/abs/hep-th/0007109}{{\ttfamily arXiv:hep-th/0007109}}.

\bibitem{Bellucci:2003yx}
S.~Bellucci, P.-Y. Casteill, and A.~Nersessian, ``{Four-dimensional Hall
  mechanics as a particle on CP**3},''
  \href{http://dx.doi.org/10.1016/j.physletb.2003.09.008}{{\em Phys. Lett.}
  {\bfseries B574} (2003) 121--128},
\href{http://arxiv.org/abs/hep-th/0306277}{{\ttfamily arXiv:hep-th/0306277}}.

\bibitem{Krivonostalk}
S.~Krivonos, ``{$HP^n$} $\sigma$-model and instanton.''
  \href{http://theorphyslab.ysu.am/sis12/Presentations/Krivonos.pdf}{Talk at
  SIS '12}, Yerevan, Armenia, 2012.

\bibitem{Hatsuda:2009vj}
M.~Hatsuda and S.~Tomizawa, ``{Coset for Hopf fibration and Squashing},''
  \href{http://dx.doi.org/10.1088/0264-9381/26/22/225007}{{\em Class. Quant.
  Grav.} {\bfseries 26} (2009) 225007},
\href{http://arxiv.org/abs/0906.1025}{{\ttfamily arXiv:0906.1025 [hep-th]}}.





\end{thebibliography}

\bibliographystyle{utphys}

\end{document}